\documentclass[10pt]{scrartcl}
\usepackage[margin=2.5cm
]{geometry}
\usepackage[utf8]{inputenc}
\usepackage[english]{babel}

\usepackage{graphicx}

\usepackage{natbib}
\usepackage{multirow, makecell}
\usepackage{enumitem}
\usepackage[normalem]{ulem}
\useunder{\uline}{\ul}{}
\usepackage{array}
\newcounter{rowcount}
\newcounter{checklistItem}
\usepackage{array} 
\usepackage{url}
\usepackage{float}
\usepackage[caption = false]{subfig}

\title{Practitioner--generated blog posts as evidence for software engineering research: attitudinal survey and preliminary checklist}
\author{
    \small{Austen Rainer}\\
    \small{Queens University Belfast} \\
    \small{Belfast, UK} \\
    \small{a.rainer@qub.ac.uk}
    \and
    \small{Ashley Williams\footnote{Corresponding Author}}\\
    \small{Manchester Metropolitan University} \\
    \small{Manchester, UK} \\
    \small{ashley.williams@mmu.ac.uk}
}

\date{June 2020}

\begin{document}

\maketitle

\begin{abstract}
Background: Blog posts are frequently used by software practitioners to share information about their practice. Blog posts therefore provide a potential source of evidence for software engineering (SE) research. The use of blog posts as evidence for research appears contentious amongst some SE researchers. There is also the significant challenge of assuring the quality of blog posts for such research.

Objective: To better understand the actual and perceived value of blog posts as evidence for SE research, and to develop guidance for SE researchers on the use of blog posts as evidence.

Method: We further analyse responses from a previously conducted attitudinal survey of 44 software engineering researchers. We conduct a heatmap analysis, simple statistical analysis, and a thematic analysis.

Results: We find no clear consensus from respondents on researchers' attitudes to the credibility of blog posts, or on a standard set of criteria to evaluate blog--post credibility. We show that \textit{some} of the responses to the survey exhibit characteristics similar to the content of blog posts, e.g., asserting prior beliefs as claims, with no citations and little supporting rationale. We illustrate our insights with \url{~}60 qualitative examples from the survey (\url{~}40\% of the total responses). We complement our quantitative and qualitative analyses with preliminary checklists to guide SE researchers in their use of blog posts as evidence.

Conclusion: Blog posts are relevant to research because they are written by software practitioners describing their practice and experience. But evaluating the credibility of blog posts, so as to select the higher--quality content, remains an ongoing challenge. The quantitative and qualitative results, together with the proposed checklists, are intended to stimulate reflection and action in the research community on the role of blog posts as evidence in software engineering research. Finally, our findings on researchers' attitudes to blog posts also provide more general insights into researchers' values for SE research.
\end{abstract}

\textbf{Keywords:} survey, blog posts, grey literature,  software practice, data quality

\newpage
\tableofcontents
\newpage
\listoffigures
\listoftables

\clearpage

\section{Introduction}
\label{section:introduction}

\subsection{Context and motivation}

The emergence of social media and of the \textit{social programmer} \citep{storey2010impact,storey2014r} has dramatically changed the way that software practitioners share information about their practice. Blog posts are one type of social media and therefore provide a potential source of evidence. Blog posts have, however, been used only sporadically as evidence in software engineering (SE) research e.g., \citep{aniche2018modern,parnin2013blogging,parnin2011measuring,pagano2011developers}.

There are significant challenges to using blog posts in research \citep{rainerASWEC2018,rainer2019using}. We consider two, related challenges in this paper. First, there is the challenge of quality--assuring blog posts to help ensure that research uses higher--quality evidence. For example, Garousi \textit{et al}. (\citeyear{garousi2018guidelines}) and Soldani \textit{et al}. (\citeyear{soldani2018pains}) both developed checklists for the quality--assurance of grey literature, however neither checklist was developed specifically for blog posts. Second, informal communication --- for example, at conferences and via reviewers' comments on our submitted manuscripts --- suggests that the use of blog posts in research (and, more generally, the use of grey literature) is contentious for at least some researchers within the community. A research community comprises a diversity of perspectives, values and goals. There is therefore the subtler challenge of understanding the research community's diversity of attitudes to blog posts as evidence for research.

In a previous paper \citep{williamsEASE19credibilitysurvey}, a conference paper, we reported initial results of a survey we conducted to study SE researchers' attitudes to the credibility of blog posts, and to criteria for evaluating such credibility. We observed in that paper that respondents adopt a conditional evaluation of the credibility of blog posts: essentially, it depends. The survey responses --- there were approximately 150 qualitative comments from 37 of the 44 respondents --- are clearly complex and rich in detail. The responses, as well as the two challenges we identify above, all warrant more attention than we were able to direct in the conference paper. We therefore further analyse the survey responses and report that analyses in this paper. We complement the analyses with checklists synthesised from the analyses of the survey and from prior literature.

\subsection{Research Questions}

For the current paper, we investigate the following research questions:
\begin{enumerate}
    \item To what degree do researchers consider blog posts to be credible? We explored this question in our conference paper \citep{rainerASWEC2018} and re--consider it here in more detail.

    \item What criteria do researchers claim to use when evaluating the credibility of a blog post? We consider both the pre--defined criteria identified \textit{a priori} for the survey, and the additional criteria that are either implied or explicitly proposed by survey respondents.

    \item What guidance on the quality--assurance of blog posts can be synthesised from prior research and the survey responses? Such guidance is helpful for establishing a standard with which researchers can evaluate blog posts.
    
    \item How do the criteria that the respondents \textit{say} they use compare with the responding \textit{behaviour} of those respondents in the survey itself? For example, a respondent may identify the lack of citations in blog posts as an indicator that blog posts lack credibility, but then the respondent may themselves not cite research in their survey responses.
\end{enumerate}

\subsection{Contributions}

The current paper substantially extends our previous analyses to make five contributions. First, we present the heatmap analysis of the quantitative responses from the survey, together with statistics complementing that heatmap. The heatmap and complementary statistics provide a much richer representation of the diversity of researchers' attitudes than the simple, and simplifying, descriptive statistics we report in the first paper \citep{williamsEASE19credibilitysurvey}. Second, we present a thematic analysis of all the qualitative responses. Our first paper simply reported a small selection of illustrative qualitative responses. Third, we present a rich sample of responses (approximately 40\% of the dataset). One reason we present that rich sample is transparency to our analysis. But the main reason we present the sample is to encourage the reader to reflect on their own attitudes to blog posts as evidence in SE research. Fourth, and as a complement to our third contribution, we show that respondents' `behaviour' in the survey (i.e., the content of some of the survey responses) has similarities with software practitioners' behaviour in their blog posts, e.g., both respondents and blog--post writers make claims, present limited reasoning, few examples, and no citations. In other words, we show that at least some threats to the credibility of blog posts are also threats to more traditional methods of data collection, such as surveys and interviews. As with our rich sample, we use this comparison of behaviours to encourage SE readers to reflect on their attitudes. Fifth, and finally, we synthesise the quantitative and qualitative analyses, together with insights from prior work, into a set of preliminary checklists that researchers can use to help them evaluate blog posts.

Overall, our analyses --- first reported in \citep{williamsEASE19credibilitysurvey} and now substantially extended in the current paper --- provides the first empirical investigation of software engineering researchers' attitudes to the value of blog posts for software engineering research, and provides the first specific guidance within the SE research community for evaluating the credibility of blog posts.

% We intend for these contributions to encourage the research community to both reflect on the actual and potential evidential value of blog posts and, more generally, of grey literature; but also to reflect on the evidential value of traditional sources of evidence.

\subsection{Structure of this paper}

The remainder of the paper is structured as follows. Section \ref{section:related-work} considers relevant, previous work, including our previous paper \citep{williamsEASE19credibilitysurvey}. Section \ref{section:survey-design} presents and discusses the design of the survey. Section \ref{section:summary-ease2019-results} briefly re--reports and summarises the quantitative results from our previous paper so as to provide context for the additional analyses reported in the current paper. Section \ref{section:quantitative-results} presents the heatmap analysis of the quantitative responses, together with statistics complementing that heatmap. Section \ref{section:qualitative-results} presents the thematic analysis of the qualitative responses. Section \ref{section:checklist} proposes checklists for evaluating facets of blog posts. The section then uses the checklists to evaluate the survey responses, comparing those with blog posts. Finally, section \ref{section:discussion} discusses our findings, considers threats to validity, identifies several directions for further research, and briefly concludes.

\section{Related work}
\label{section:related-work}

\subsection{The use of blog posts in SE research}

Table \ref{table:examples-of-research-using-blogs} presents examples of empirical research into software practice that make use of blog posts. As the table indicates, blog posts have been used in software engineering research for about a decade, albeit sporadically, with the quantity of blog posts used in the research ranging from one blog post \citep{rainer2017using} to 50,000 blog posts \citep{pagano2011developers}.

The examples presented in the table are organised into three broad categories: 1) \textit{primary studies} that use blog posts to gain insights into software practice; 2) \textit{secondary studies} that use grey literature, such as blog posts, to gain insights into software practice; and 3) \textit{guidelines} papers that present advice on the use of grey literature (including blog posts) in research, these guidelines including quality checklists. 

Our focus in this paper is on the use of blog posts as evidence for software engineering research. We recognise that there are other fields of research that use blog posts. These fields include opinion mining, sentiment analyses, argumentation mining and experience mining, e.g., \citep{swanson2014identifying,gordon2009identifying, burton2009icwsm,park2010detecting,kurashima2009discovering,inui2008experience,kurashima2006mining}. Other research, e.g., \citep{lakshmanan2010knowledge, chau2009blog, khan2017modelling}, is also interested in blog posts but not necessarily empirically, or with the intent to study software practice.

The three categories identified in Table \ref{table:examples-of-research-using-blogs} approach the credibility of blog posts in contrasting ways. Secondary studies and guidelines develop and apply quality checklists for assessing the quality of grey literature, with Garousi \textit{et al}. (\citeyear{garousi2018guidelines}) proposing a more generic checklist as part of their guidelines for Multivocal Literature Reviews (MLRs)\footnote{There are important differences between the systematic review of previously published primary studies and the grey literature review of blog posts.}. Primary studies of software practice tend to accept the quality of selected blog posts without evaluation (e.g., there is no post--selection quality evaluation) however in their respective primary studies Williams (\citeyear{williams2018using}) used reasoning indicators as a proxy for quality, and Rainer (\citeyear{rainer2017using}) selected a blog post from the blog of one highly regarded practitioner.

\begin{table}[ht]
    \small
    \center
    \caption{Research into software practice using blogs}
    \label{table:examples-of-research-using-blogs}
    %\centering
    \begin{tabular}{  p{4.6cm}  l  p{5.6cm} }
    \hline\noalign{\smallskip}
    \textbf{First author} & \textbf{Year} & \textbf{\# blog posts}\\
    \hline\noalign{\smallskip}
    %\hline\noalign{\smallskip}
    \multicolumn{3}{ l }{\textit{Guidelines on secondary studies of GL in research}}\\
    % \hline\noalign{\smallskip}
    \citep{garousi2018guidelines} & 2018 & N/A\\
    % \hline\noalign{\smallskip}
    %\hline\noalign{\smallskip}
    \multicolumn{3}{ l }{}\\
    \multicolumn{3}{ l }{\textit{Secondary studies that use blog posts in research}}\\
    % \hline\noalign{\smallskip}
    \citep{soldani2018pains} & 2018 & 20/51 blog posts (40\% of dataset) \\
    %\hline\noalign{\smallskip}
    \citep{RaulamoJurvanen2017choosing} & 2017 & 60 GL sources\\
    %\hline\noalign{\smallskip}
    \citep{garousi2016and} & 2016 & 46 internet articles \& white papers\\
    % \hline\noalign{\smallskip}
    %\hline\noalign{\smallskip}
    \multicolumn{3}{ l }{}\\
    \multicolumn{3}{ l }{\textit{Primary studies that use blog posts in research}}\\
    % \hline\noalign{\smallskip}
    \citep{williams2018software} & 2018 & 2852 blogs posts\\
    %\hline\noalign{\smallskip} 
    \citep{rainer2017using} & 2017 & One blog post\\
    %\hline\noalign{\smallskip}
    \citep{parnin2013blogging} & 2013 & 300 blog posts\\
    %\hline\noalign{\smallskip}
    \citep{parnin2011measuring} & 2011 & 376 blog posts\\
    %\hline\noalign{\smallskip}
    \citep{pagano2011developers} & 2011 & 50,000 blog posts\\
    \hline\noalign{\smallskip}
    \end{tabular}
\end{table}

\subsection{Blog posts and classifications of evidence}

The research community already classifies blog posts as a candidate source of evidence. We consider two examples of that recognition here: Prechelt and Petre's (\citeyear{prechelt2010credibility}) discussion of credibility, and Wohlin's (\citeyear{wohlin2013evidence}) evidence profile for software engineering research and practice.

Prechelt and Petre (\citeyear{prechelt2010credibility}) identify five sources of evidence: (direct) experience, other people, reflection, reading, and scientific (or quasi--scientific) exploration. For the category of \textit{reading}, they state, ``Written materials, both informal (such as \textit{high--quality blog posts}) or formal (such as scientific articles) transport insights from other parties.'' (emphasis added here). There are relationships between these sources of evidence. Two relationships between the sources are particularly relevant to the current paper. First, a practitioner gains \textit{experience} (we italicise the categories proposed by Prechelt and Petre), \textit{reflects} on that experience, and then shares that reflective experience as a blog post for others to \textit{read}. Second, a practitioner gains \textit{experience}, \textit{reflects} on that experience, and then shares that reflective experience with researchers as part of a \textit{scientific exploration} that is then published for others to \textit{read}. Prechelt and Petre (\citeyear{prechelt2010credibility}) highlight an important qualifier, of course: the reading of \textit{high--quality blog posts}. This is precisely the category of blog post that our work (in the current paper and elsewhere) aims to identify.

Wohlin (\citeyear{wohlin2013evidence}) developed an evidence profile to be used to organise different types of evidence for a particular case. Wohlin identifies seven sources of information, from the strongest source to the weakest, and discounts the three weakest sources as not evidence. Of the remaining four sources, two are relevant here: statements from trustworthy witnesses, and statements from expert witnesses. Blog posts may contain information from trustworthy witnesses and expert witnesses. As with Prechelt and Petre (\citeyear{prechelt2010credibility}), there is an important qualifier: identifying those blog posts that are written by \textit{trustworthy} witnesses and expert witnesses.

\subsection{Standards of evidence}

Prechelt and Petre (\citeyear{prechelt2010credibility}) also recognise that ``Different purposes require different \textit{standards of evidence}.'' (\citep{prechelt2010credibility}; emphasis in original). \citep{devanbu2016belief} and \citep{rainer2003persuading} both show that practitioners tend to hold a different standard of evidence to researchers: practitioners prefer the opinion of other practitioners as their primary source of evidence. Our interest in the current paper is of course on researchers' standards.

A further complication is that researchers hold different worldviews. Petersen and Gencel (\citeyear{petersen2013worldviews}) consider the relationship between worldviews, research methods and validity. They assert that the dominant worldview in empirical software engineering is the pragmatist worldview, in which multiple research methods from the other worldviews (i.e., positivist, constructivist, and participative) are used. For Petersen and Gencel, the multi--method approach creates difficulties in reporting threats to validity, and creates confusion when evaluating research. Petersen and Gencel's work highlights the challenges of defining a standard or standards for evidence. Our analysis later in this paper provides insights into the diversity of attitudes of researchers. Also, in developing a suite of checklists, we contribute toward the establishment of a standard (or standards) for evaluating blog posts as evidence.

Aside from the challenges that arise from different worldviews, there is a broad distinction between the relevance of evidence, the validity of that evidence, and the rigour of the process by which that evidence has been generated. We draw on two prior publications to consider relevance, validity and rigour: Prechelt and Petre (\citeyear{prechelt2010credibility})'s book chapter on credibility, and Ivarsson and Gorschek's (\citeyear{Ivarsson2011}) evaluation of the industrial relevance and rigour of technology evaluations in software engineering.

Prechelt and Petre (\citeyear{prechelt2010credibility} broadly distinguish between relevance and credibility (for Prechelt and Petre, credibility comprises high validity and good reporting). An implication from Prechelt and Petre (\citeyear{prechelt2010credibility})'s position on relevance and credibility is that an evaluator would typically consider relevance first, and then go on to consider the rigour of relevant claims and evidence. It follows that blog posts might be relevant to a topic of interest to a researcher, for which these blog posts are \textit{then} evaluated for their validity of claims, and possibly for the rigour by which the claims were formed. 

Ivarsson and Gorschek (\citeyear{Ivarsson2011}) develop a model for evaluating the industrial relevance and rigour of technology evaluations in software engineering. They define methodological rigour in terms of \textit{conformance} to standards for that methodology i.e., a study is methodologically rigorous if that study conforms to guidelines, protocols and best practices for that methodology. One way in which researchers quality--assure evidence published in research papers is by checking the rigour of the process by which the evidence was generated. One significant difficulty with blog posts is that there is no standard, \textit{de facto} or otherwise, for the writing of a blog post, or for the enquiry process that informs the claims then stated in the blog post, or indeed for the reporting of that enquiry process.

% Ivarsson and Gorschek's (\citeyear{Ivarsson2011}) definition of rigour is similar in intent to Liebchen and Shepperd's (\citeyear{liebchen2008data, liebchen2016data}) focus on data quality. Liebchen and Shepperd define data quality in terms of accuracy (minimisation of noise). Ivarsson and Gorschek (\citeyear{Ivarsson2011}) and Liebchen and Shepperd (\citeyear{liebchen2008data, liebchen2016data}) recognise other dimensions to validity and quality e.g., \textit{completeness} in the degree of conformance to standards, or of a dataset; and \textit{fitness for purpose}. What constitutes fitness for purpose by definition depends on the evaluator's purpose, and by implication on their worldview, so that fitness for purpose has an element of subjectivity to it.

% \subsection{Evidence}

% Pfleeger and Menezes (\citeyear{pfleeger2000marketing}) discuss how evidence can be used by practitioners to make decisions on technology adoption. They consider how practitioners first assess the credibility of evidence, and then combine multiple pieces of credible evidence in order to form conclusions. Credibility assessment differs between researchers and practitioners and, for Pfleeger and Menezes, researchers should produce evidence that is useful, and therefore relevant, to practitioners and credible (rigourous) to practitioners and researchers.

\subsection{Quality--assurance of publications}

% The research discussed in the preceding subsections provides broad principles or perspectives on credibility and on evidence. There is the need to support such principles with specific guidelines and checklists.

Over many years, the empirical software engineering community has reflected on how to quality--assure publications.
% presented questions for evaluators to use to evaluate the credibility of claims and evidence.
For example, twenty--five years ago, Fenton \textit{et al}. (\citeyear{fenton1994science}) proposed several questions to ask about any claim arising from software engineering research.
% : 1) Is the claim based on empirical evaluation and data? 2) Is the experiment (study) designed correctly? 3) Is the claim based on (the study of) a toy or real situation? 4) Were the measurements used appropriate to the goals of the experiment (study)? 5) Was the experiment (study) run for a long enough time?
And in a founding paper on Evidence Based Software Engineering (EBSE) Dyb{\aa} \textit{et al}. (\citeyear{dyba2005evidence}) proposed five questions to help practitioners assess the credibility of evidence: 1) Is there any vested interest? 2) Is the evidence valid?, 3) Is the evidence important? 4) Can the evidence be used in practice? 5) Is the evidence consistent with evidence from other studies?

More recently, questions asking about claims and evidence have been structured into checklists. Two checklists are particularly relevant for the  current paper because both relate to grey literature: Garousi \textit{et al}.'s (\citeyear{garousi2018guidelines}) generic checklist for MLRs, and Soldani \textit{et al}.'s (\citeyear{soldani2018pains}) specific checklist for a GLR of microservices. We summarise these checklists here in Table \ref{table:quality-checklists}. Later in this paper we integrate these checklists  with the quantitative and qualitative results from our survey.

\begin{table*}[!htbp]
    \small
    \centering
\caption{Quality checklists.}
\label{table:quality-checklists}
\begin{tabular}{ p{0.3cm}  p{14.7cm} }
\hline\noalign{\smallskip}
& \textbf{Question}\\ 
\hline\noalign{\smallskip}
%\hline
\multicolumn{2}{ l }{\textbf{Garousi \textit{et al}.'s (\citeyear{garousi2018guidelines}) quality checklist. }}\\
\multicolumn{2}{ l }{\textit{Authority of the producer}}\\
\hline\noalign{\smallskip}
& Is the publishing organization reputable?\\
%\hline
& Is an individual author associated with a reputable organization?\\
%\hline
& Has the author published other work in the field? \\
%\hline
& Does the author have expertise in the area?\\
\hline\noalign{\smallskip}
%\hline
\multicolumn{2}{ l }{\textit{Methodology}}\\
\hline\noalign{\smallskip}
& Does the source have a clearly stated aim? \\
%\hline
& Does the source have a stated methodology?\\
%\hline
& Is the source supported by authoritative, contemporary references?\\
%\hline
& Are any limits clearly stated? \\
%\hline
& Does the work cover a specific question? \\
%\hline
& Does the work refer to a particular population or case?\\
%\hline
\hline\noalign{\smallskip}
\multicolumn{2}{ l }{\textit{Objectivity}}\\
\hline\noalign{\smallskip}
& Does the work seem to be balanced in presentation?\\
%\hline
& Is the statement in the sources as objective as possible? Or, is the statement a subjective opinion?\\
%\hline
& Is there vested interest?\\
%\hline
& Are the conclusions supported by the data?\\
%\hline
\hline\noalign{\smallskip}
\multicolumn{2}{ l }{\textit{Date}}\\
\hline\noalign{\smallskip}
& Does the item have a clearly stated date?\\
\hline\noalign{\smallskip}
%\hline
\multicolumn{2}{ l }{\textit{Position regarding related resources}}\\
\hline\noalign{\smallskip}
& Have key related grey literature or formal sources been linked to / discussed?\\
\hline\noalign{\smallskip}
%\hline
\multicolumn{2}{ l }{\textit{Novelty}}\\
\hline\noalign{\smallskip}
& Does it enrich or add something unique to the research?\\
%\hline
& Does it strengthen or refute a current position?\\
%\hline
\hline\noalign{\smallskip}
\multicolumn{2}{ l }{\textit{Impact}: A normalisation of several impact metrics}\\
\hline\noalign{\smallskip}
& Number of citations, backlinks, media shares, comments, views\\
%\hline
\hline\noalign{\smallskip}
\multicolumn{2}{ l }{\textit{Outlet type} (see \citep{garousi2018guidelines})}\\
\hline\noalign{\smallskip}
& 1st Tier, 2nd Tier, 3rd Tier \\
\hline\noalign{\smallskip}
\multicolumn{2}{ l }{Soldani \textit{et al}.'s (\citeyear{soldani2018pains}) quality checklist.}\\
\hline\noalign{\smallskip}
\textbf{\#} & \textbf{Explanation} \\ 
\hline\noalign{\smallskip}
%\hline
\multicolumn{2}{ l }{Inclusion criteria}\\
\hline\noalign{\smallskip}
I1 & The study discusses the industrial application of microservices.\\
%\hline
I2 & The study discusses the benefits or shortcomings of microservice design, development or operation.\\
%\hline
I3 & The study reports on direct experiences, opinions or practices on microservices by educated practitioners.\\
%\hline
I4 & The study refers to a practical case--study of design, development or operation of microservices.\\
%\hline
\hline\noalign{\smallskip}
\multicolumn{2}{ l }{Exclusion criteria}\\
\hline\noalign{\smallskip}
E1 & The study does not offer details on design or implementation of microservices.\\
%\hline
E2 & The study is not referred to industrial cases or other factual evidence.\\
%\hline
E3 & The benefits or pitfalls of microservices are not justified/quantified by the study.\\
%\hline
E4 & The study does not provide scope and limitations of proposed solutions/patterns. \\
%\hline
E5 & The study does not offer evidence of a practitioner perspective.\\
\hline\noalign{\smallskip}
%\hline
\multicolumn{2}{ l }{Additional control factors}\\
\hline\noalign{\smallskip}
C1 & Practical experience: A study is to be selected only if it is written by practitioners with 5+ experience in service-oriented design, development and operation, or if it refers to established microservices solutions with 2+ years of operation. \\
%\hline
C2 & 
Industrial case-study: A study is to be selected only if it refers to at least 1 industrial case-study where a quantifiable number of microservices are operated.\\
%\hline
C3 & Heterogeneity: The selected studies reflect at least 5 top industrial domains and markets where microservices were successfully applied.\\
%\hline
C4 & Implementation quantity: The selected studies refer to/show implementation details for the benefits and pitfall they discuss, so that other researchers and practitioners can use them in action.\\
\hline\noalign{\smallskip}
\end{tabular}
\end{table*}

As well as checklists, there are a range of guidelines available, including: Kitchenham and Charters' (\citeyear{Kitchenham07guidelinesforSLRs}) guidelines for conducting systematic reviews; Petersen \textit{et al}.'s (\citeyear{petersen2015guidelines}) guidelines for systematic mapping studies; Garousi \textit{et al}.'s (\citeyear{garousi2018guidelines}) guidelines for MLRs; Runeson and H{\"o}st (\citeyear{runeson2009guidelines}) guidelines and checklists for case study research (see also \citep{runeson2012case}); Kitchenham and Pfleeger's (\citeyear{kitchenham2002surveypart5}) guidelines for survey research; and Kitchenham \textit{et al}.'s (\citeyear{kitchenham2002preliminary}) general guidelines on empirical research. Whilst all of these questions, checklists and guidelines assert, or prescribe, how to evaluate information, none of these guidelines are particular to blog posts. 

\subsection{Summary}
Blog posts are already being used in SE research and blog posts are already recognised within existing classifications of evidence. There are concerns about the credibility of blog posts and these concerns broadly relate to the validity of the statements made in blog posts as well as the rigour of the processes of enquiry (if any) that inform those statements. Prior to a consideration of validity or of rigour, the evaluator might (typically) first consider the relevance of the blog post, i.e., whether the content of the blog post is relevant to the research topic of interest to the researcher. In principle, many blog posts would be relevant to research because those blog posts refer to aspects of software practice. There are, legitimately, different standards to evidence. We are of course interested here in the standards of evidence as they concern researchers, and therefore the standards that researcher do or ought to place on blog posts as evidence for research. Finally, different worldviews influence what the respective researchers consider acceptable standards of evidence and therefore influence what any given researcher would accept as a relevant and valid blog post for research.

\section{Survey Design}
\label{section:survey-design}

\subsection{Our previous study}

Given the current status of software engineering research that uses blog posts, we decided to conduct an investigation into researchers' attitudes to the credibility of blog posts. We discuss the design of our survey later in this section. For the current subsection, we briefly present a critical review of the first paper \citep{williamsEASE19credibilitysurvey} we published that reported the preliminary results of the survey.

Our initial analysis \citep{williamsEASE19credibilitysurvey} suggested a complex relationship between researchers' general attitude to the credibility of blog posts and the researchers' credibility criteria for evaluating blog posts. As one example, and as we show later in this paper, it appears that \textit{some} researchers' place a greater value on criteria such as reasoning and professional experience, with relatively less value placed on reporting empirical data and methods of data collection. By contrast, other researchers placed greater value on reporting empirical data and methods of data collection. These contrasting positions appear consistent with Prechelt and Petre's (\citeyear{prechelt2010credibility}) assertion about different standards of evidence, and Petersen and Gencel's (\citeyear{petersen2013worldviews}) arguments concerning the effect of worldview on the assessment of validity.

% assertion that the research community's goal is ``\dots\space systematic, well--informed, evidence--based \textit{critical thinking}.'' (\citep{prechelt2010credibility}; emphasis in original). 

It became clear, as we continued to analyse the data, that there were several limitations to our survey design. For example, the dataset is relatively small (as is the relative size of the software engineering research community), there were a limited number of questions in our survey (though this was deliberate to encourage responses to the survey) and the nature and range of permissible responses (because of the Likert scaling) limited the ability to explore researchers' attitudes. We did however allow for qualitative responses.

For the current paper, we report a more detailed analyses of the data. We report quantitative analyses, comprising a heatmap and statistical analysis (whilst recognising that meaningful \textit{statistical} analyses of the data may be both difficult and potentially misleading). We complement the quantitative analyses with a thematic analysis of the qualitative responses.

Initial results for the survey, together with a full description of the survey design, were first reported in \citep{williamsEASE19credibilitysurvey}. For completeness, we restate in this section the description of the survey design. We complement that description with a brief explanation of our heatmap analysis of the quantitative data and our thematic analysis of the qualitative data. In section \ref{section:summary-ease2019-results} we briefly re--report the results from the first paper.

\subsection{Candidate criteria}
\label{subsection:candidate-criteria}
To inform the design of our survey, we conducted a broad review of previous empirical studies of credibility. Many of these studies were published outside of software engineering research. We identified 833 candidate articles for review, reducing that candidate set to a final set of 13 papers. The full details of the literature review are reported in a technical report \citep{williams_rainer_cred_tr}.

From the set of 13 papers, we identified 88 candidate criteria, which we subsequently distilled to the nine criteria we used in the survey. These nine criteria are summarised in Table \ref{table:summary-of-nine-criteria}. Respondents to the survey identified additional criteria which we discuss later in this paper.

\begin{table}[!htpb]
    \small
    \center
    \caption{Summary of the nine criteria from the literature review}
    \label{table:summary-of-nine-criteria}
    \begin{tabular}{ | p{7cm} | p{1.3cm} | }
    \hline
    \textbf{Criterion} & \textbf{Acronym}\\
    \hline
    Clarity of writing & CoW\\
    Reporting empirical data & RED \\
    Reporting the method of data collection & RM\\
    Reporting professional experience & PExp\\
    Reasoning & Rsn\\
    Citing practitioner sources & URL--P\\
    Citing research sources & URL--R\\
    Prior beliefs of the reader & PB\\
    Prior beliefs of others who influence the reader & IoO\\
    \hline
    \end{tabular}
\end{table}

\subsection{Survey development and refinement}
\label{subsection:survey-refinement}
We developed a draft survey that was reviewed by four colleagues who were familiar with the research, and revised the survey in response to their feedback. We then conducted a pilot study, inviting responses from a network of software engineering researchers within New Zealand (SI\^{}NZ\footnote{http://softwareinnovation.nz/}). Based on the feedback from the pilot study we clarified some survey questions. The final survey comprised 20 questions, comprising 12 main questions of which one question had nine sub--questions. The questions and permissible answers are summarised in the Appendix.

We used the Qualtrics\footnote{https://qualtrics.com} online survey instrument to administer the survey. An invitation email was sent to each participant, with instructions and an anonymous link to the survey. The survey was approved by the appropriate University of Canterbury (New Zealand) Ethics Committee (HEC 2017/68/LR-PS).

The main questions in the survey asked about: 1) the researcher's opinion on the general credibility of blog articles, 2) the criteria they (say they) use to assess the credibility of blog articles, 3) whether there are any criteria that we have not identified; 4) whether they think the criteria generalise to other grey literature, 5) whether they think the criteria generalise to research, and 6) whether they had any other comments to make on the survey. Each of these questions included the opportunity for the respondent to provide open--ended, additional comments. It is these comments that form the bases for the thematic analysis.

As optional questions, we also asked respondents for their contact details, whether they would be willing to participate in a follow--up interview, and whether they would like to receive an anonymised copy of the data. 21 respondents (47\%) confirmed they would like a copy of the data, and the anonymised raw data has been sent to those respondents. The full list of questions is available online at \url{https://www.researchgate.net/publication/324784268_Design_of_a_survey_on_credibility}.

\subsection{Participants}

The survey was conducted between 13\textsuperscript{th} February 2018 and 26\textsuperscript{th} March 2018. Invitations were sent out to the Programme Committees of the \textit{Evaluation and Assessment in Software Engineering (EASE)} conference and the \textit{Empirical Software Engineering and Measurement (ESEM)} conference for that year. `Overlapping' members for each Programme Committee were emailed only once, and respondents involved in the development and refinement of the survey were not invited to participate in the full survey.

Overall, 138 researchers were invited to participate. Four of these invitees asked us whether they could forward the survey to their colleagues. We approved these requests but were not able to track increase in numbers of invited participants. Consequently, we are unable to precisely report the number of people who actually received the invitation. 57 invitees started the survey and 44 completed it, giving a response rate of 32\% (assuming 138 invitees). The participants' experience in research ranged from two years to 35 years, with a mean average of 16.2 years. 

% A summary of respondents' research interests is given in Table \ref{table:respondents-research-interests}.

%\begin{table}[htbp]
% \begin{table}[ht]
%     \centering
%     \small
%     \caption{Summary of respondents' research interests}
%     \label{table:respondents-research-interests}
%     \begin{tabular}{| l | r |  l | r | } 
%         \hline
%         \textbf{Interest} & \textbf{\textit{f}} & \textbf{Interest} & \textbf{\textit{f}} \\ 
%         \hline
%         \hline
%         mining \& analytics 	&	17	&	        evolution 	&	2	\\
%         testing 	&	15	&	        global s/w development 	&	2	\\
%         empirical SE 	&	14	&	        open source 	&	2	\\
%         human factors 	&	14	&	        project management 	&	2	\\
%         other 	&	13	&	        security 	&	2	\\
%         requirements engineering 	&	12	&	        software product 	&	2	\\
%         quality 	&	10	&	        technical debt  	&	2	\\
%         software processes 	&	9	&	        usability  	&	2	\\
%         agile 	&	8	&	        behavioral s/w engineering 	&	1	\\
%         research 	&	6	&	        programming 	&	1	\\
%         metrics 	&	5	&	        risk 	&	1	\\
%         software engineering 	&	4	&	        safety 	&	1	\\
%         EBSE 	&	3	&	        startups 	&	1	\\
%         maintenance 	&	3	&		&		\\
%         \hline
%     \end{tabular}
% \end{table}

We intentionally designed a short survey to encourage participation. We estimated, from our draft and pilot study, that the survey would take about 10 minutes to complete. The total time taken to complete the survey ranged from 2.4 minutes to 22 hours with an overall average of 75.7 minutes. Ignoring the completion times of the five responses that took longer than one hour to complete gives a range from 2.4 minutes to 47.1 minutes, with an average time of 11.7 minutes. 

\subsection{Post--survey follow--up}

We performed two stages of post--survey follow--up. For the first stage, at the close of the survey, we contacted invitees to ask them for reasons for why she or he did not start the survey, or started the survey but did not complete it. As the survey was anonymous, we emailed all invitees as we could not email only those who had not completed the study. A similar follow--up was also conducted with the SI\^{}NZ trial. The reasons given with both survey--trial follow--up and the full survey follow--up was that respondents were too busy to start, or complete, the survey. 

For the second stage, upon completion of the initial analysis and acceptance of the first paper \citep{williamsEASE19credibilitysurvey}, we sent a copy of the results to those respondents who had expressed an interest in receiving the results. As noted earlier, 21 respondents (47\%) were sent a copy of the data.

\subsection{Excluded--case analyses}
\label{subsection:outlier-analyses}

During the quantitative analysis for our first paper \citep{williamsEASE19credibilitysurvey}, we identified one respondent --- respondent \#22 in Table \ref{figure:heatmap-of-responses} --- who scored 0 (zero; no blog post is credible) to the question asking about the general credibility of blog posts. The same respondent also provided maximum scores of 1 for two of the criteria, Reporting empirical data and Reporting research methods, with scores of 0 for all other criteria. This is a surprising (even confusing) set of quantitative responses because respondent \#22 appears to be stating that criteria such as Reasoning and citations have no value at all (a score of 0) and that, with scores of 1, Reporting empirical data and Reporting research methods have minimal importance. 

In her or his qualitative responses, respondent \#22 states, ``It is simply impossible to evaluate the value [of blogs] since no real evidence is provided.'', and ``If we start trusting blogs, we might as well stop doing scientific research in software engineering.'' The survey was not asking whether blog posts reported empirical evidence or reported research methods but rather was asking what the respondent considered to be important criteria for \textit{evaluating} the trustworthiness of blog posts. Given respondent \#22's qualitative comments, we might expect their scores on the credibility criteria to be high, i.e., at least high scores for Reporting empirical data and Reporting research methods, perhaps with at least relatively high scores for Reasoning and citations to research. Yet, as noted above, these are not the scores that were reported. It seems that respondent \@22 has misinterpreted the questions that were asked in the survey. Given the unusual responses from respondent \#22, and also to remain consistent with our previous paper \citep{williamsEASE19credibilitysurvey}, we remove case \#22 from our quantitative analyses. We do however retain the respondent's comments for our thematic analysis. 

A second respondent --- \#39 in Table \ref{figure:heatmap-of-responses} --- also demonstrates unusual responding behaviour. This respondent completed the survey in 2.4m, the quickest time to complete the survey, scored 3 for the General Credibility question, and then scored 5 consistently for all nine credibility criteria, returning one of the highest total scores across all 44 respondents. Again, this response is surprising. The respondent completed the survey about 5 times quicker than the adjusted average for all respondents and reports a uniform set of responses across all nine criteria. The respondent makes few qualitative comments, the only comment being ``It [the general quality of blog posts] depends. Quality and credibility of blog articles vary wildly.''. For our quantitative and thematic analysis we retain case \#39.

In section \ref{section:qualitative-results} we look again at possible excluded cases when we consider the qualitative responses.
% For the analyses reported in the current paper, we \textit{retain} all cases because we are primarily interested in the variety and richness of the respondents' qualitative responses.
% We identify cases that could be considered for exclusion. Such cases may be outliers rather than inaccurate (`noisy') data (cf. \citep{liebchen2016data}).

\subsection{Quantitative analysis}

Although our dataset is relatively small (n=44) with a limited number of questions and a limited range of values (typically 0 -- 6), the attitudinal data is complex and we don't want to over--simplify the data by only using descriptive statistics. We also don't want to mislead: whilst a statistical test might be more sophisticated it does not necessarily produce a more reliable insight. We therefore report a heatmap of all 44 responses and complement that heatmap with statistical analysis. We use RStudio (v1.2.5033) with the \texttt{tidyverse} (v1.3.0) collection of R packages.

\subsection{Thematic analyses}
% \subsubsection{Overview}
To complement each quantitative question in the survey we invited the respondent to provide open--ended comments. A total of 143 comments were received from the 44 respondents.

We first analysed separately the qualitative comments associated with each quantitative question. We read several times each set of comments, and iteratively developed \textit{nodes} to denote respondents' concepts, discussed the nodes, then developed themes and discussed the themes.

For the current paper, we report thematic analyses of all comments as one dataset. We present verbatim quotes from a relatively large proportion (approximately 40\%) of the comments.  We do this to convey the richness of the opinions held by the 44 respondents. We used NVivo for Mac (v11.4.2) to qualitatively analysis the data. 

% We do this because our primary interest is to better understand the perspectives of the respondents. Asking different questions of the respondents is a helpful way of gathering information from respondents, but isolating analyses to each question can constrain the analyses and the resulting findings. Indeed, as discussed in section \ref{section:models-and-evaluation}, we have developed a conception of the evaluation of the credibility of blog posts that is much more complex compared to the conception we commenced with, and treating the qualitative data as one whole dataset has helped with that transformation of conception.

We choose not to report frequencies of coded qualitative data for several reasons. First, our motivation for reporting qualitative comments is to provide a rich complement to the quantitative analyses. Quantifying the qualitative data dilutes that richness and `just' provides another set of quantitative results. Second, a respondent can provide qualitative comments for each and all of the criteria. Given that we analyse the dataset as a whole, there could be a `double counting' of frequencies of qualitative codes (where a code repeats across comments because the comment is repeated by a respondent). As one illustrative example, we found some respondents proposing additional criteria that were already present in the survey. As another illustrative example, we found cases were respondents were making comments such as, `Please refer to my previous comments' which raises complications in objectively counting nodes. Third, the quantity of comments declined as the respondents progressed through the survey, with reduced variation and frequency of codes identified for later criteria. This may lead to an imbalanced dataset, were one to compare comments across different questions.

\subsection{Public access to data}

For transparency, the raw responses from all respondents are available online\footnote{https://www.researchgate.net/publication/331704210}.

\section{Summary of previous results}
\label{section:summary-ease2019-results}

As noted, we first reported initial results from the survey in a conference paper \citep{williamsEASE19credibilitysurvey}. We seek a balance of providing sufficient foundation and background from the previous conference paper, so as to support the new quantitative and thematic analyses reported in the current paper, whilst not just re--reporting here the results of that previous paper. We re--report from \citep{williamsEASE19credibilitysurvey} our analyses of the respondents' overall attitude to the credibility of blog posts (see section \ref{subsection:general-credibility} and \ref{fig:general-credibility}) to provide necessary context for the current paper, and then summarise the findings from the previous paper.

\subsection{The general credibility of blog posts}
\label{subsection:general-credibility}
\begin{figure}[htbp]
    \centering
    \includegraphics[width=0.8\textwidth]{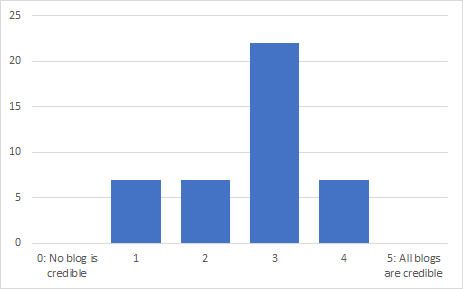}
    \caption{The general credibility of practitioner blog posts (n=43, Mod=3, Med=3, Mn=2.7)}
    \label{fig:general-credibility}
\end{figure}

The survey asked respondents whether, in general, they consider blog posts to be credible. Figure \ref{fig:general-credibility} presents the results. We are conscious that respondents are being asked to evaluate a complex situation (i.e., a very large volume of blog posts that vary in content and quality) with a one--score response. Given the complexity of the situation, a score of 3 might constitute the `safest' response, or the most conservative response, for many respondents.

\subsection{Summary of results from initial analysis}
\label{subsection:summary-initial-results}

We conducted the following analyses, with  observations and conclusions in our conference paper \citep{williamsEASE19credibilitysurvey}:
\begin{enumerate}
    \item We ranked the nine credibility criteria according to the 7--point Likert scale. We found
    \begin{enumerate}
        \item A surprisingly low level of rankings, e.g., 32\% of respondents consider the \textit{Reporting of empirical data} to be an extremely important criteria. We speculated that these low percentages may indicate that respondents do not consider criteria to be essential, or perhaps the respondents were making some kind of adjustment in their valuation for blog posts.
        \item The Reasoning criteria was most frequently ranked as extremely important (~40\%).
        \item Devanbu \textit{et al}. \citeyear{devanbu2016belief} and Rainer \textit{et al}. \citeyear{rainer2003persuading} found that software engineering practitioners valued their own personal experience, and that of their colleagues, over independent, third--party empirical evidence. By contrast, researchers assign a low importance to \textit{Prior beliefs} and \textit{Influence of others}.
    \end{enumerate}
    
    \item We separated the data into three subsamples based on the degree to which respondents considered blog posts credible: \textit{Low} credibility (scores of 0 and 1), \textit{Medium} credibility (scores of 2 and 3) and \textit{High} credibility (scores of 4 and 5. We found:
    \begin{enumerate}
        \item The \textit{Low} subsample ranked four criteria --- \textit{Clarity of writing} (CoW), \textit{Reporting empirical data} (RED), \textit{Reporting methods of data collection} (RM) and \textit{Reasoning} (Reason) --- relatively highly. These four criteria most clearly align with the overt values of the empirical software engineering research community.
        
        \item The \textit{Low} subsample had a minimum value of 5 for \textit{Reporting empirical data} (RED). This could suggest that respondents in the \textit{Low} subsample place a particularly high value on reporting empirical data. This might be a spurious result in our data, but as was noted it is a result consistent with the overt values of an \textit{empirical} research community. 
        
        \item 71\% of respondents in the \textit{High} subsample rated the \textit{Reason} criterion as extremely important. This contrasted with 43\% and 31\% in the \textit{Low} and \textit{Medium} subsamples retrospectively, and suggests that respondents in the \textit{High} subsample place a particularly high value on reasoning. This observation might explain why this subsample considers blog posts to generally be credible: the respondents in this subsample primarily evaluate credibility in terms of the presence of reasoning, and blog posts provide a flexible medium for practitioners to express their reasoning. A related observation is that none of the respondents in the \textit{High} subsample consider the \textit{Reporting methods of data collection} (RM), citations to practitioner sources (URL-P), or citations to research sources (URL-R) to be extremely important.
        
        \item The contrasting views \textit{between} the three subsamples together with the contrasting views \textit{within} each subsample may help to explain why blog posts are contentious. As one example, the \textit{High--credibility} respondents place a high value on \textit{Reasoning} and a low value on \textit{Reporting empirical data} for blog posts, in contrast to the \textit{Low--credibility} respondents who place a relatively high value on \textit{Reporting empirical data}.
    \end{enumerate}
    
    \item We also asked respondents whether they thought the identified nine criteria generalised to other practitioner--generated content and to research content. We found that:
    \begin{enumerate}
        \item Over 60\% of respondents thought that the criteria generalise to other practitioner--generated content.
        \item A very similar percentage, over 58\%, thought the criteria also apply to researcher--generated content.
    \end{enumerate}
\end{enumerate}

\section{Exploratory analyses of the quantitative data}
\label{section:quantitative-results}

\subsection{Summary of responses}

\begin{figure}[htbp]
    \centering
    \includegraphics[width=0.8\textwidth]{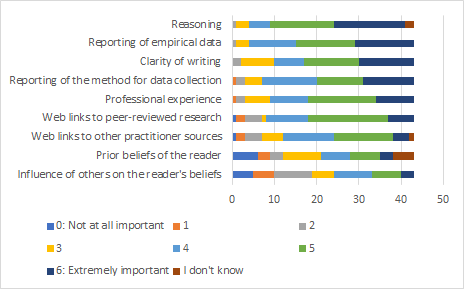}
    \caption{Stacked barchart of responses to the nine criteria}
    \label{fig:stacked-barchart-general-credibility}
\end{figure}

Figure \ref{fig:stacked-barchart-general-credibility} presents a stacked barchart of respondents' ratings of the nine credibility criteria. We report this stacked barchart to provide a context for the subsequent analyses.

\subsection{Heatmap analysis}
\label{section:heatmap-analysis}

\begin{figure}[htpb]
\centering
\includegraphics[scale=0.6]{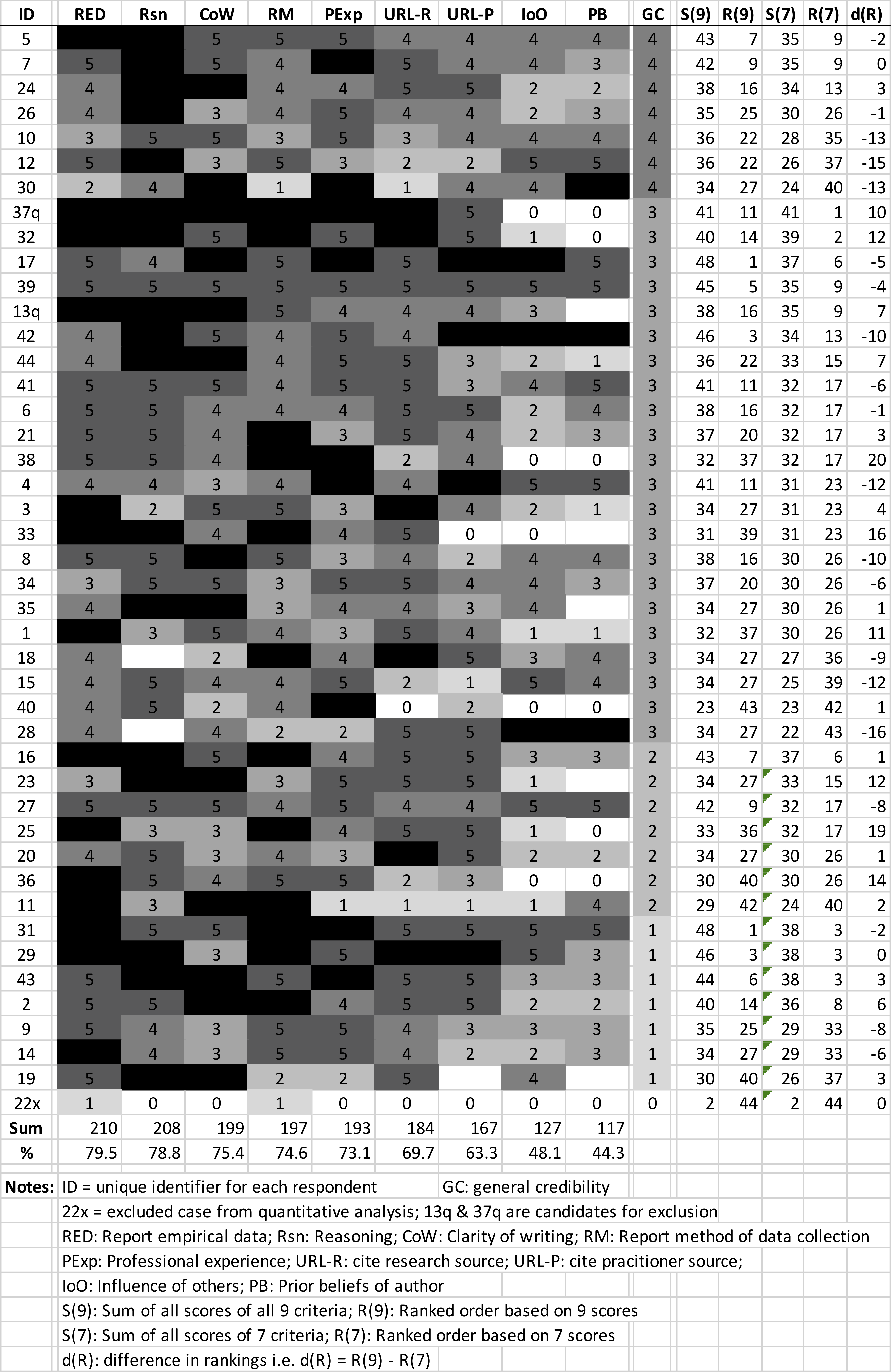}
\caption[Heatmap of responses]{Heatmap of quantitative responses (n=44)}
\label{figure:heatmap-of-responses}
\end{figure}

Figure \ref{figure:heatmap-of-responses} presents a heatmap of the respondents' scores for the nine credibility criteria and the general credibility of blog posts, together with descriptive statistics for those scores.

The heatmap is ordered using the following `logic'. The heatmap is first ordered from left to right based on the sum of \textit{all} respondents' scores for each criteria (see the penultimate row at the foot of the heatmap, labelled \textbf{Sum}). The heatmap is then ranked from top to bottom, first based on the respondent's score for general credibility (see the column labelled \textbf{GC}) and then based on \textit{seven} of the credibility criteria (see the column labelled \textbf{R(7)}). The seven criteria used in the ranking are the first seven ordered in the figure i.e. RED, Rsn, CoW, RM, PExp, URL--R and URL--P. Our previous paper \citep{rainerASWEC2018} found that the last two criteria, IoO and PB, were considered to be qualitatively different to the first seven criteria. We discuss this difference below.

Recall that scores for each of the credibility criteria are in the range 0\footnote{Does a score of 0 mean ``I don't know''? Check.} to 6, whilst scores for the general credibility of blog posts are in the range 0 to 5 (though no respondent provided a score of 5). Cells shaded entirely black have the score of 6. All other cells contain the respective respondent's score. Cells not containing a value represent responses of ``I don't know'' or missing values. Respondents that are excluded from the quantitative analysis are indicated with an \url{x} suffix against the ID, i.e., 22\url{x}. Respondents that are candidates for exclusion, as identified during the subsequent qualitative analysis, are indicated with an \url{q} suffix against the ID, i.e., 13\url{q}, 37\url{q}. Subsection \ref{subsection:outlier-analyses} discusses excluding cases on the basis of the quantitative analysis and subsection \ref{subsection:candidate-cases-for-exclusion} discusses excluding cases on the basis of the qualitative analysis.

The logic by which the heatmap has been ordered helps to highlight a general pattern of attitudes amongst the respondents. Respondents tend to show a preference for the criteria toward the left side of the heatmap, and they show this preference regardless of how they rate the credibility of blog posts. 

\subsection{Actual response proportions relative to theoretical range}

The \textit{theoretical} range for an \textit{individual} respondent's sum of scores for all nine credibility criteria is 0 to 54 (nine credibility criteria each with a range of scores from 0 to 6), with a theoretical median average score for a respondent of 27.5 (as there are \textit{55} values [0 - 54] available in the sum of scores). Similarly, the theoretical range for the sum of scores for an individual \textit{criterion}, across all 44 respondents, is 0 to 264 (44 respondents with a maximum score of 6; this range obviously reduces as one excludes cases). 

% \textcolor{red}{For ease of interpretation, Figure \ref{figure:line-chart-of-criteria-totals} visualises the final two rows of the heatmap, i.e., shows a line chart of sum of scores for each of the criteria, together with the percentage of the theoretical maximum score. (Note that the \textit{x}--axis for Figure \ref{figure:line-chart-of-criteria-totals} has two scales, an absolute scale and a percentage scale.) Figure \ref{figure:line-chart-of-criteria-totals} therefore provides a more abstract summary of the heatmap, again showing a preference for some of the criteria.}

% \begin{figure}[htpb]
% \centering
% \includegraphics[scale=0.5]{images/pseudo_scree_plot_for_criteria.png}
% \caption[Line chart of criteria totals]{Line chart showing sum of scores for each of the criteria, together with percentage of theoretical maximum score (n=44)}
% \label{figure:line-chart-of-criteria-totals}
% \end{figure}

\subsection{Confidence intervals}

Table \ref{table:confidence-intervals} reports 95\% confidence intervals\footnote{Calculated using the following guidance: \url{http://www.stat.yale.edu/Courses/1997-98/101/catinf.htm}.} for the percentage responses to the nine credibility criteria\footnote{We have not done a pairwise exclusion of cases to compensate for missing responses later in the survey.}. In other words, given the proportions reported in the \textit{sample}, the confidence intervals provide a range estimate for the \textit{population}'s proportion for each criteria. Confidence intervals are calculated using the \textit{z}--score of a Normal population, of course, so the estimates reported in Table \ref{table:confidence-intervals} obviously assume that the population of survey responses for each criterion would conform to a Normal distribution. This is unlikely to be a reasonable assumption, however reporting confidence intervals helps us to explore the data.

(To digress briefly: one candidate explanation for a non--Normal distribution is that there may be distinct sub--populations, e.g., a population of software engineering researchers who have not and do not read blog posts, in contrast to a population of software engineering researchers who have and do read blog posts. This explanation is based on the qualitative comments relating to self--excluding cases, as discussed in subsection \ref{subsection:candidate-cases-for-exclusion}. Each of these sub--populations will have different experiences of the value and credibility of blog posts and therefore their attitudes will likely contrast.)

Accepting the difficulties with the confidence intervals, the table suggests that the \textit{upper limits} to the confidence intervals for the Others and Beliefs criteria are very close to the \textit{lower limits} for the top four criteria. This suggests that variation in the \textit{sample means} -- perhaps arising because of the convenience sampling -- would be \textit{unlikely} to affect a number of the results. Acknowledging the difficulties with the confidence intervals, the best we can conclude at this stages is that the broad `profile' of proportions in the sample \textit{may} be present in the population.

\begin{table}[!ht]
    \caption{Confidence intervals for proportions of responses for the nine criteria}
    \label{table:confidence-intervals}
    % \small
    \centering
    \begin{tabular}{| p{1.7cm} | p{1.2cm} | p{1.8cm} | p{1.8cm} | p{1.8cm} | }
    \hline
  \textbf{Criteria}	&	\textbf{Count}	&	\textbf{Proportion}	&	\textbf{CI Upper}	&	\textbf{CI Lower}	\\
    \hline
    \hline
    RED	&	210	&	0.795	&	0.915	&	0.676	\\
    Rsn	&	208	&	0.788	&	0.909	&	0.667	\\
    CoW	&	199	&	0.754	&	0.881	&	0.626	\\
    RM	&	196	&	0.742	&	0.872	&	0.613	\\
    Pexp	&	193	&	0.731	&	0.862	&	0.600	\\
    URLR	&	184	&	0.697	&	0.833	&	0.561	\\
    URLP	&	167	&	0.633	&	0.775	&	0.490	\\
    Others	&	127	&	0.481	&	0.629	&	0.333	\\
    Beliefs	&	117	&	0.443	&	0.590	&	0.296	\\
    \hline
    \end{tabular}
\end{table}

% Recognising the qualifiers with the confidence intervals, there are at least two contrasting interpretations of the data:
% \begin{enumerate}
%     \item If one accepts as reasonable the assumption of a Normal distribution of responses, then the table indicates that the \textit{upper limits} to the confidence intervals for the Others and Beliefs criteria are very close to the \textit{lower limits} for the top four criteria. This suggests that variation in the \textit{sample means} -- perhaps arising because of the convenience sampling -- would be \textit{unlikely} to affect a number of the results. In other words, the broad `profile' of proportions in the sample are likely to be present in the population.
    
%     \item If one does not accept as reasonable the assumption of a Normal distribution of responses, then an alternative interpretation is that the data is suggesting at least two different sub--populations.
% \end{enumerate}

\subsection{Correlation analysis}

Table \ref{table:spearman-correlations} presents the Spearman Rho correlations for each pair of the nine criteria. The correlations may be grouped as follows:

\begin{enumerate}
    \item positive correlations that have a high Rho value and are statistically significant (p $<$ 0.05);
    \item positive correlations that have a relatively high Rho value but are not statistically significant (p $>=$ 0.05);
    \item positive correlations that have a relatively low Rho value;
    \item correlations that have a negative Rho value (negative correlations);
    \item those correlations that have a very low, or even no, correlation
\end{enumerate}

\begin{table}[!ht]
    \caption{Spearman Rho rank correlations for the nine credibility criteria}
    \label{table:spearman-correlations}
    \small
    \centering
    \begin{tabular}{| p{1.5cm} | r | r | r | r | r | r | r | r | }
        \hline
        \textbf{Criterion} & \textbf{RED} & \textbf{Rsn} & \textbf{CoW} & \textbf{RM} & \textbf{Pexp} & \textbf{URLR} & \textbf{URLP} & \textbf{Others}\\
        \hline
        \hline
        % RED	&	1	&		&		&		&		&		&		&		\\
        Rsn	&	-0.07	&		&		&		&		&		&		&		\\
        CoW	&	-0.01	&	0.23	&		&		&		&		&		&		\\
        RM	&	0.72	&	-0.04	&	-0.16	&		&		&		&		&		\\
        Pexp	&	-0.15	&	0.05	&	0.00	&	-0.03	&		&		&		&		\\
        URLR	&	0.21	&	0.19	&	0.06	&	0.24	&	-0.17	&		&		&	\\
        URLP	&	-0.06	&	0.12	&	0.06	&	0.09	&	0.19	&	0.56	&		&\\
        Others	&	-0.29	&	0.05	&	0.06	&	-0.32	&	0.10	&	-0.08	&	0.25	&		\\
        Beliefs	&	-0.33	&	-0.06	&	0.09	&	-0.33	&	0.02	&	-0.24	&	0.11	&	0.85\\

        \hline
    \end{tabular}
\end{table}

As indicated by the groupings, most correlations are not statistically significant. This may partly be explained by the low sample size (and hence low statistically power). We focus here on the two correlations that are significant, and on the negative correlations.

\begin{figure}
    \centering
    \subfloat[Beliefs \& influence] {\includegraphics[scale=0.3]{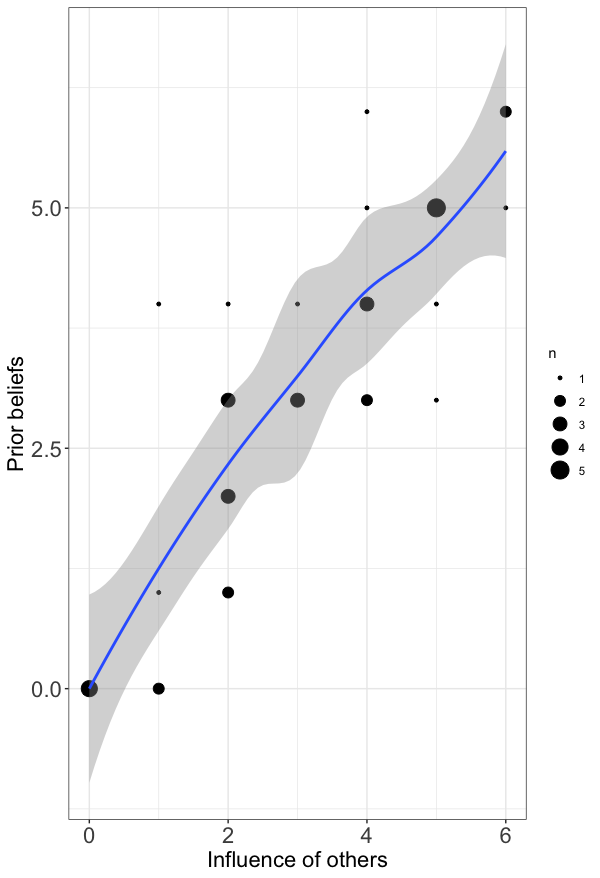}\label{subfigure:Beliefs-Others-correlation}}
    \subfloat[Data \& method]{\includegraphics[scale=0.3]{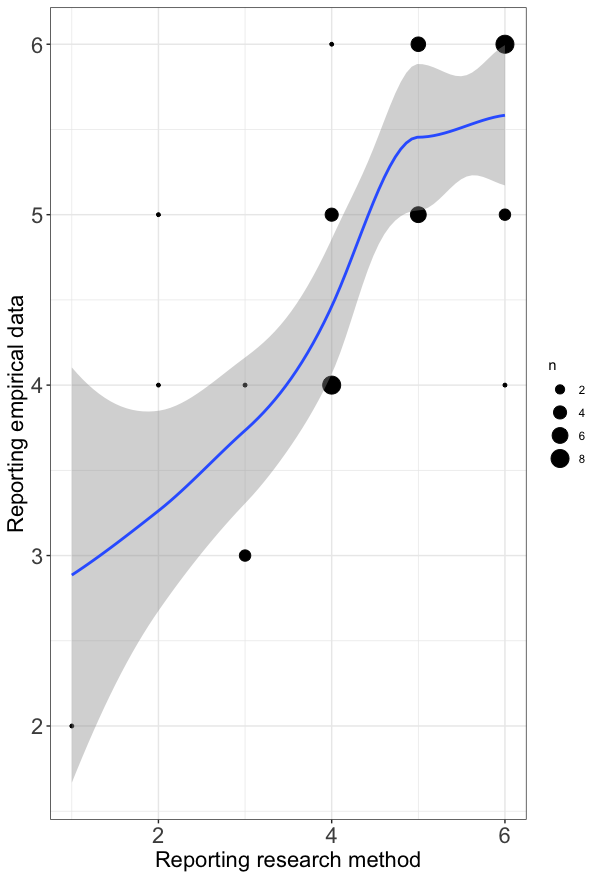}\label{subfigure:RED-RM-correlation}}\\
    \subfloat[Influence \& Methods] {\includegraphics[scale=0.3]{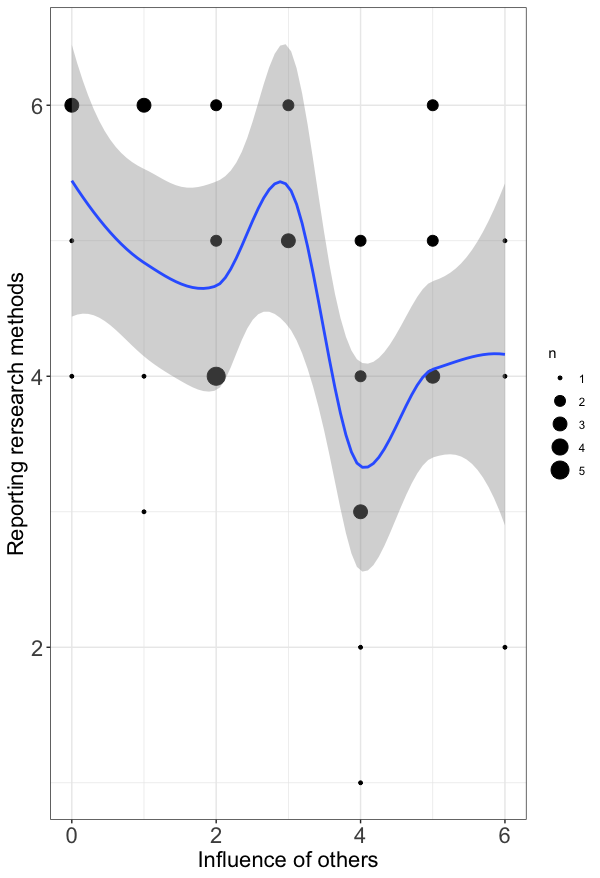} \label{subfigure:Others-RM-correlation}}
    \subfloat[Data \& Beliefs]{\includegraphics[scale=0.3]{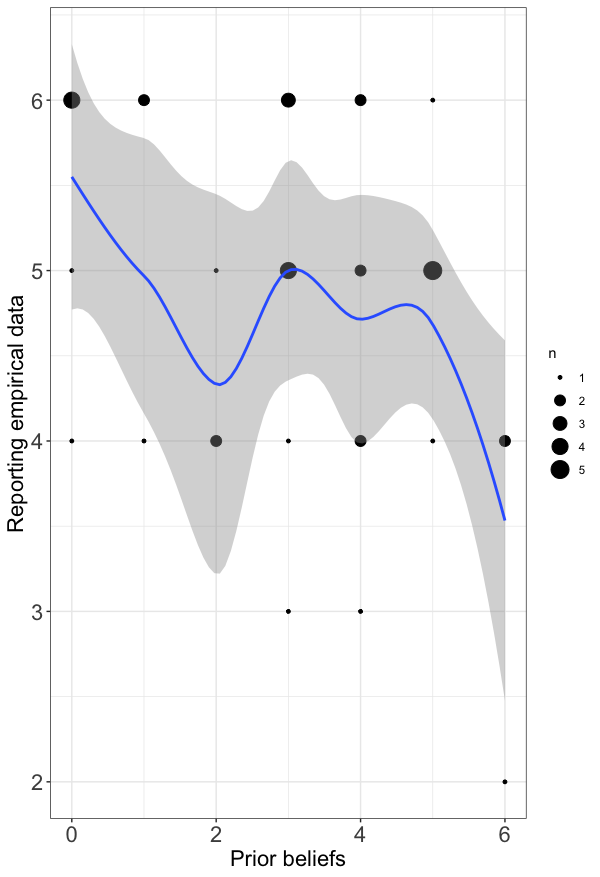}\label{subfigure:RED-Beliefs-correlation}}\\
    \caption{Bubble charts with Loess smoother and confidence interval}
    \label{figure:interesting-correlations}
\end{figure}

\begin{enumerate}

    \item There is a very strong statistical correlation between the Influence of others (Others) criterion and the Prior beliefs (Beliefs) criterion (Rho = 0.85). The correlation is visualised in Figure \ref{subfigure:Beliefs-Others-correlation}. These two criteria are likely to be qualitatively different to the other seven criteria because these two criteria are not directly `measurable' within the blog post itself. Our heatmap also suggests these two criteria are different. We think that these two criteria and their correlation would be `caused' by an underlying effect that is \textit{different} to the other seven criteria. We return to this pair of criteria in due course.
    
    \item There is a strong correlation between the Reporting empirical data (RED) criterion and the Reporting research method (RM) criterion (Rho = 0.72). The correlation is visualised in Figure \ref{subfigure:RED-RM-correlation}. That there is a correlation, and that this correlation is one of the stronger correlations, is not surprising. We expect researchers to independently rank both criteria highly (or relatively highly) because the research community places high importance on evidence and on the rigour of evidence--collection. Given the manifest importance of these criteria to the community, and given the strength of their correlation, these two criteria and their collection could act as a kind of benchmark against which to consider the other criteria and correlations. For example, earlier (see subsection \ref{section:heatmap-analysis}) we speculated on the possibility of different sub--populations. The strength of correlation between the Reporting empirical data (RED) criterion and the Reporting research method (RM) criterion suggests two criteria that are relatively constant across sub--populations.
    
\end{enumerate}

We also observe \textit{negative} correlations between the pair of criteria, Reporting empirical data and Reporting research methods, and the pair of criteria, Prior beliefs and the Influence of others. Two examples of the negative correlations are shown in Figures \ref{subfigure:Others-RM-correlation} and \ref{subfigure:RED-Beliefs-correlation} for illustration. These correlations are not statistically significant, however they are consistent with the values of the software engineering research community, i.e., to value evidence over one's prior beliefs and over the influence of others.

Bubble plots for each of the criteria can be found in the Appendix. The plots present a kind of transposition of the data presented in the heatmap.

% \subsection{Relatively low correlations}

% There are quite low correlations between several pairs of criteria:

%     \begin{enumerate}
%         \item Reporting empirical data (RED) and citing prior research (URLR);
%         \item Reporting research methods (RM) and citing prior research (URLR);
%         \item Reasoning (Rsn) and citing prior research (URLR);
%         \item Reasoning and Clarity of writing (CoW);
%         \item Professional experience and citing practitioners' publications (URLP).
%     \end{enumerate} 

\subsection{Summary}

% Table \ref{table:summary_of_quantitative_analysis} presents a summary of the quantitative analyses.
We have examined the quantitative data from various perspectives in the search for patterns within the data that would provide insights into respondents' attitudes to the use of blog posts in research. At best, our overall analyses provide the following limited insights:
\begin{enumerate}

    \item The sample is relatively small and imbalanced, and this effects the ability to quantitatively analyse the data, and the confidence we can draw from that analysis.

    \item There appears to be a general preference for certain credibility criteria, i.e. Reporting empirical data, Reasoning, Clarity of writing, Reporting research methods and Professional experience.  These preferences appear to be regardless of the rating for General Credibility. 
    
    \item Although there is a quantitative ranking to these preferred criteria, the closeness of the rankings and the confidence intervals suggest the ordering of the sample may not reflect the ordering in the population (but the confidence intervals are likely untrustworthy because of the nature of the data).
    
    \item There are two pairs of criteria that have within--pair correlations. These are:
    \begin{enumerate}
        \item Reporting empirical data correlating with Reporting research methods; and
        \item Influence of others correlating with Prior beliefs.
    \end{enumerate}
    
    \item The correlation of Reporting empirical data with Reporting research methods is not unsurprising, given the nature of the research discipline, i.e., with importance placed on evidence and evidence--collection.
    
    \item The correlation of Influence of others correlating with Prior beliefs may be due to a misunderstanding of these criteria, or because these two criteria are qualitatively different to the seven other criteria.

    \item There \textit{may} be a negative correlation between to the two pairs of criteria, i.e., Reporting empirical data and Reporting research methods negatively correlating with Influence of others and Prior beliefs. Again this is not unsurprising, given the nature of the research discipline, i.e., with importance placed on evidence and evidence--collection.
    
    \item There may be sub--populations in the data.
    
\end{enumerate}

% \begin{table}[!ht]
%     \caption{Summary of quantitative analyses}
%     \label{table:summary_of_quantitative_analysis}
%     \centering
%     \begin{tabular}{| p{1.5cm} | c | p{1.5cm} | p{6.5cm} |}
%         \hline
%         \textbf{Criterion} & \textbf{Rank} & \textbf{Heatmap total} & \textbf{Significant correlations} \\
%         % \hline
%         \hline
%         RED & 1 & 210 & Strong correlation with RM\\
%         Rsn & 2 & 208 & \\
%         CoW & 3 & 199 & \\
%         RM  & 4 & 197 & Strong correlation with RED\\
%         Pexp & 5 & 193 & \\
%         URLR & 6 & 184 & \\
%         URLP & 7 & 167 & \\
%         Others & 8 & 127 & Very strong correlation with Beliefs\\
%         Beliefs & 9 & 117 & Very strong correlation with Influence of others\\
%         \hline
%     \end{tabular}
% \end{table}
\section{Thematic analyses}
\label{section:qualitative-results}

\subsection{Introduction}

Our qualitative analyses identified a number of themes. In the following subsections, we present tables of verbatim quotes from the respondents and discuss those quotes. To aid cross--referencing, each quote has a unique identifier of the format \texttt{Q\#}. Inevitably, some quotes refer to more than one theme. To save space, in most cases we present the quote in one table but embolden appropriate phrases to emphasise the themes as well as the richness of the data. For each quote, we include the respondent's rating of the general credibility of blog posts, of the format \texttt{GC=n}, where \texttt{n} is the rating in the range 0 (lowest) to 5 (highest). For each table, we order the quotes according to the respondents' scores for General Credibility.

\subsection{General comments on respondents and responses}
% \subsection{Candidate cases for exclusion and inclusion}
\label{subsection:candidate-cases-for-exclusion}

Table \ref{table:general-observations} presents examples of three issues relating to respondents and their responses: examples where respondents and their responses might be excluded; examples of where respondents are aware of variability in the quality of blog posts; and examples of respondents' expectations of blog posts.

Table \ref{table:general-observations}a) identifies examples of respondents who could be excluded/included from the analyses, depending on how strict one sets the exclusion/inclusion criteria. Examples for exclusion could be treated as either outliers or inaccurate (noisy) cases. Respondent \texttt{Q\ref{quote:respondent-22}} is the respondent already identified as an outlier from our earlier analyses \citep{williamsEASE19credibilitysurvey}. Respondent \texttt{Q\ref{quote:don't-follow-blogs}} bases her or his assessment of the general credibility of blog posts (GC=3) on her or his ``impression''.  Similarly, \texttt{Q\ref{quote:quality-varies}} is unusual as it is one of the few respondents in the survey who confirm they read a lot of blog posts.

% \subsection{The recognised variability of blog posts}

Table \ref{table:general-observations}b) indicates that survey respondents are aware of variability in the quality of blog posts and therefore, by implication, of the difficulty of responding.

% \subsection{Expectations of the respondents}

Table \ref{table:general-observations}c) provides comments on the respondents' expectations of blog posts. Expectations may also be understood as a kind of prior belief, and may give rise to biases, such as anchoring bias or confirmation bias. For example, if I expect blog posts to lack credibility I may not be alert to those blog posts that provide credible information. 

We did not explicitly ask respondents to state their expectations. One interpretation of our results is that, in fact, the results constitute a baseline of researchers' expectations of blog posts.

%%% AUSTEN - start of tables
\begin{table}[!htpb]
    \small
    \center
    \caption{General observations on respondents and responses}
    % \label{table:other-outliers}
    \label{table:general-observations}
    \begin{tabular}{ p{0.3cm}  p{14.7cm} } 
        \hline\noalign{\smallskip}
        \textbf{\texttt{Q\#}} & \textbf{Comment}\\
        \hline\noalign{\smallskip}
        \multicolumn{2}{l}{\textbf{\ref{table:general-observations}a) Example cases for exclusion and inclusion}}\\
        \hline\noalign{\smallskip}
        \multicolumn{2}{l}{\textit{Candidate cases for exclusion}}\\
        \refstepcounter{rowcount} \arabic{rowcount} \label{quote:do-not-read-many} & This is just a feeling. \textbf{I do not read many blog articles.} [GC=3]\\
        \refstepcounter{rowcount} \therowcount \label{quote:don't-follow-blogs} & \textbf{I do not follow blogs}. My comments relate to \textbf{my impression} of the work of practitioners such as Don Reifer, Capers Jones, Larry Putnam, Dan Galorath; organizations such as NESMA and IFPUG; and data providers such as ISBSG. [GC=3]\\

        \refstepcounter{rowcount} \therowcount \label{quote:respondent-22} & It is simply impossible to evaluate the value since no real evidence is provided. [GC=0]\\

        \noalign{\smallskip}
        \multicolumn{2}{l}{\textit{Candidate cases for inclusion}}\\

        \refstepcounter{rowcount} \arabic{rowcount} \label{quote:read-some} & \textbf{I like to read some blog articles} to get a closer look at what practitioners are doing. [GC=2]\\
        
        \refstepcounter{rowcount} \therowcount \label{quote:quality-varies}& \textbf{I do read a lot of blogs}. The quality varies. [GC=3]\\

        \noalign{\smallskip}
        \multicolumn{2}{l}{\textit{Other cases}}\\
        
        \refstepcounter{rowcount} \therowcount \label{quote:never-used-blogs}& \textbf{I have never used blog info in my research}, rather use empirical data collected directly from companies. Pls, consider this when analysing my answers! [GC=1]\\
        \hline\noalign{\smallskip}

        \multicolumn{2}{l}{\textbf{\ref{table:general-observations}b) Recognition of the variability of blog posts}}\\
        \hline\noalign{\smallskip}

        \refstepcounter{rowcount} \arabic{rowcount} \label{quote:empty1} & I think we should assess the type of information according to its [the author's] \textbf{intention}. It is unfair to assess a blog post as we would do to a scientific paper, and vice--versa. They have \textbf{different target audience}. I understand that blog posts that \textbf{present some empirical data} or \textbf{method} [of data collection] have additional positive characteristics, while scientific works that present neither of them have additional negative characteristics. Both can be assessed regarding a their \textbf{reasoning} and the use of \textbf{practical experiences}, though. [GC=3]\\

        \refstepcounter{rowcount} \arabic{rowcount} \label{quote:empty2}  & You can go from a really good, well-backed article to a completely non-sense, personal feeling-based article [\textbf{bias}]. Since anyone can write anything, there should be a way to filter it, recommend, or something like this. [GC=3]\\

        \refstepcounter{rowcount} \arabic{rowcount} \label{quote:impartiality-1} & It really depends. Some are sh*t and some make a lot of sense. [GC=3]\\

        \refstepcounter{rowcount} \arabic{rowcount} \label{quote:quality-varies-again} & I do read a lot of blogs. The quality varies. [GC=3]\\

        \refstepcounter{rowcount} \arabic{rowcount} \label{quote:vary-wildly-2} & It depends. Quality and credibility of blog articles vary wildly. [GC=3]\\
        \hline\noalign{\smallskip}

        \multicolumn{2}{l}{\textbf{\ref{table:general-observations}c) Expectations of blog posts (grouped by GC score)}}\\
        \hline\noalign{\smallskip}
        \refstepcounter{rowcount} \therowcount \label{quote:empty3} & I dont (sic) usually read a blog post from a practitioner \textbf{expecting} it to contain any of this information. If they present this type of information is a plus to me. [GC=3]\\

        \refstepcounter{rowcount} \therowcount \label{quote:empty4} & \textbf{What I expect} to find in a blog post to assess its quality is a \textbf{good reasoning} for the writers' argumentation and, when possible, some \textbf{experience report}. The \textbf{level of detail} in both the \textbf{reasoning} the writer uses and the \textbf{experience report} the writer presents is what I consider the most important to assess the post's quality. [GC=3]\\

        \refstepcounter{rowcount} \therowcount \label{quote:impartiality-1-again} & It really depends. Some are sh*t and some make a lot of sense. [GC=3]\\

        \refstepcounter{rowcount} \therowcount \label{quote:vary-wildly} & It depends. Quality and credibility of blog articles vary wildly. [GC=3]\\
        \multicolumn{2}{l}{}\\
        
        \refstepcounter{rowcount} \therowcount \label{quote:my-experience-1} & \textbf{I expect blogs} to be principally \textbf{anecdotal} - that is the \textbf{value} for me in them - do the anecdotes/war stories align with \textbf{my experience (as practitioner)} and do they align with \textbf{empirical results} or offer \textbf{counter-evidence}. I would be pleasantly surprised if they offered more than \textbf{simple empirical observations} that could be calibrated. [GC=2]\\
        
        \hline\noalign{\smallskip}
    \end{tabular}
\end{table}

\subsection{Author bias, subjectivity and sincerity}

Table \ref{table:comments-relating-to-bias-and-honesty}a) presents contrasting comments on (i) author bias and subjectivity, and (ii) author sincerity. A sincere practitioner may unintentionally be bias, a contrast that is also found with researchers e.g., as demonstrated with the need for calculating inter--rater agreement on assessments. Bias, subjectivity and sincerity are not properties that only apply to the writing of blog posts. Information shared by any practitioner, or researcher, for example through interview, survey, focus group or blog post has a degree of bias, subjectivity and sincerity. 

\subsection{Topic and content}

Table \ref{table:comments-relating-to-bias-and-honesty}b) presents comments on the content of blog posts. Quote \texttt{Q\ref{quote:credible-to-me}} suggests that researchers' evaluations of the credibility of a blog post is partly dependent on the researchers' interest in \textit{specific content} of the blog post. For example, Treude, Parnin, Storey and Aniche (e.g. \citep{parnin2011measuring, parnin2013blogging, aniche2018modern}) have a particular interest in the degree to which practitioners discuss API methods in blog posts. They therefore also have an interest in the presence of code fragments that demonstrate or illustrate the API method/s. Pagano and Maalej (\citeyear{pagano2011developers}) observed that only 1.8\% (934) of the 50,701 blog posts they examined (from four active open source projects) contained source code paragraphs. This contrasts with Parnin and Treude (\citeyear{parnin2011measuring}) who observed that 90\% of the posts they examined (336 posts from 373 posts) had code snippets in the post, a median of 3 code snippets per post. In both studies, the researchers were interested in the presence of code, but found a very different prevalence of the presence of code. This difference could affect how one interprets the credibility of a blog post or the dataset of blog posts, or indeed blog posts in general. Evaluating the credibility of a blog post in relation to the researchers' interest in \textit{specific content} connects back to Petersen and Gencel's (\citeyear{petersen2013worldviews}) work on worldviews.

Another example is Stack Overflow. The nature of the question--and--answer site is that, to the degree possible, the questioner is encouraged to ask a Minimal, Complete and Verifiable Example (MCVE)\footnote{\texttt{https://stackoverflow.com/help/mcve}; an alternative acronym is Minimal Worked Example (MWE): a collection of source code and other resources that allow the bug or problem to be demonstrated and reproduced.}: use the minimal amount of code needed to (re)produce the problem; complete all parts (of the code) to (re)produce the problem; and test the code first to verify that you can (re)produce the problem. A blog post of this kind would be different to, for example, a discussion of the relative benefits and drawbacks of different testing techniques,

The qualitative data therefore suggests the following types of content (this list is not exhaustive): empirical data; opinions (beliefs); experience; Reasoning; and source code.

% \textcolor{magenta}{Not sure if the following remains relevant: Soldani \textit{et al}. \citep{soldani2018pains} have conducted a grey literature of microservices. And Garousi \textit{et al}. \citep{garousi2016need} have conducted a multivocal literature review of automated testing. These topics might be more suited to credible information in blog posts, in contrast to for example requirements gathering, because they constitute more discrete, concrete situations.}

\begin{table}[!htpb]
    \small
    \center
    \caption{Observations on author subjectivity, blog content, and professional experience}
    \label{table:comments-relating-to-bias-and-honesty}
    \begin{tabular}{ p{0.3cm}  p{14.7cm} } 
        \hline\noalign{\smallskip}
        \textbf{\texttt{Q\#}} & \textbf{Comment}\\
        \hline\noalign{\smallskip}
        
        \multicolumn{2}{l}{\textbf{\ref{table:comments-relating-to-bias-and-honesty}a) Author bias, subjectivity and sincerity (grouped by GC score)}}\\
        \hline\noalign{\smallskip}
        
        \multicolumn{2}{ l }{(i) \textit{Comments relating to author bias and subjectivity}}\\
        \hline\noalign{\smallskip}

        \refstepcounter{rowcount} \therowcount \label{quote:empty5} & They are based on \textbf{personal opinions} or gut feeling, very often \textbf{subjective} or \textbf{inspired by someone else}. [GC=3]\\

        \refstepcounter{rowcount} \therowcount \label{quote:believe-1} & The blogs are credible in that they represent the view of the bloggers. \textbf{I don't believe they fake up their opinions}. Of course, the views may be \textbf{highly biased}. [GC=3] \\

        \refstepcounter{rowcount} \therowcount \label{quote:empty6} & In general, blog discuss ideas or advocate for technique/tools. Assesments (sic) could be \textbf{biased}. [GC=3]\\

        \refstepcounter{rowcount} \therowcount \label{quote:empty7} & You can go from a really good, well-backed article to a completely non-sense, personal feeling-based article [\textbf{bias}]. Since anyone can write anything, there should be a way to filter it, recommend, or something like this. [GC=3]\\
        \multicolumn{2}{l}{}\\
        
        \refstepcounter{rowcount} \therowcount \label{quote:believe-2} & While \textbf{I do not think that blog articles are deliberately wrong or misleading}, I believe that \textbf{they are very subjective} opinions and just should be treated accordingly (that is: cautiously). [GC=2]\\
        \multicolumn{2}{l}{}\\
        
        \refstepcounter{rowcount} \therowcount \label{quote:empty8}& It is hard to read such materials. I am always concerned with their validity and fairness. Usually they represent opinions (sometimes \textbf{biased}) on some topics. [GC=1]\\
        \multicolumn{2}{l}{}\\
        
        \hline\noalign{\smallskip}
        \multicolumn{2}{ l }{(ii) \textit{Comments relating to author sincerity}}\\
        \hline\noalign{\smallskip}

        \refstepcounter{rowcount} \therowcount \label{quote:believe-3} & Blogs provide opinions, so from the point of view of those writing the blogs, they \textbf{believe} that what they are stating is \textbf{true and reliable}. However, such assertions may fail a more detailed scrutiny. [GC=3]\\
        
        \refstepcounter{rowcount} \therowcount \label{quote:empty9} & Blog articles represent the \textbf{experience} of the writer and so are ``credible'' from that point of view.  I would be \textbf{surprised to find any blog writer deliberately misrepresenting} his or her \textbf{observations}. [GC=3]\\
        
        \refstepcounter{rowcount} \therowcount \label{quote:empty10} & It's often there own Opinon; \textbf{mostly serious, cause they don't want to blame themselves with stupid staff}; interesting views Often \textbf{experiences} that can be very useful and that are from practice. [GC=3]\\
        \multicolumn{2}{l}{}\\
        
        \refstepcounter{rowcount} \therowcount \label{quote:my-experience-2} & In \textbf{my experience} blog articles are typically written in \textbf{good faith} by practitioners who have a degree of \textbf{familiarity with their chosen topic}. Whether such individuals necessarily have \textbf{the wider context} of their chosen topic or perspective on the topic can be more open to question. [GC=2]\\
        \hline\noalign{\smallskip}

        \multicolumn{2}{l}{\textbf{\ref{table:comments-relating-to-bias-and-honesty}b) Content of the blog posts}}\\
        \hline\noalign{\smallskip}
        %\hline
        % \multicolumn{2}{| l |}{(a) Comments relating to topic}\\
        % \hline
        \refstepcounter{rowcount} \therowcount \label{quote:credible-to-me} & I mostly think it depends on the \textbf{subject matter} of the post. Generally \textbf{blogs and Q\&A sites on programming are credible to me} and quite useful as well. On the other hand, when a blog post regards other fields of software engineering such as \textbf{management, requirements, measurement and even testing}, it is quite hard to identify a post that has zero or little \textbf{bias}. When they try to make some comparison with alternatives to their recommendation, it is usually based on \textbf{outdated information} or unfair comparison. [GC=3]\\
        %\hline
        \refstepcounter{rowcount} \therowcount \label{quote:empty11} & It depends on the \textbf{topic} as well as on the \textbf{amount} of provided content and its \textbf{level of detail}. [GC=3]\\
        %\hline
        \refstepcounter{rowcount} \therowcount \label{quote:empty12} & It depends on \textbf{what they are reporting}. [GC=3]\\
        \hline\noalign{\smallskip}

        \multicolumn{2}{l}{\textbf{\ref{table:comments-relating-to-bias-and-honesty}c) Professional experience (grouped by GC score)}}\\
        \hline\noalign{\smallskip}
        %\hline
        \refstepcounter{rowcount} \therowcount \label{quote:empty13}  & (In the field of agile development) \textbf{practitioners experiences} are very important. The credibility of course depends on the quality of the \textbf{content}. It also depends on the format, would probably also trust a blog article more than a single short answer in a Q\&A section. [GC=4]\\
       \multicolumn{2}{l}{}\\
       
        \refstepcounter{rowcount} \therowcount \label{quote:empty14}  & The main goal of blogs articles is \textbf{not to provide empirical evidence}, each blogger has his/her own goal from \textbf{sharing experiences}, self-marketing, getting public attention, etc. [GC=2]\\

        \refstepcounter{rowcount} \therowcount \label{quote:empty15}  & \textbf{Experience} that are grounded in prior research, supports or refutes prior finding, explains the \textbf{context} and \textbf{data collection method} clearly seems credible to me. [GC=2]\\
        \multicolumn{2}{l}{}\\
        
        \refstepcounter{rowcount} \therowcount \label{quote:empty16}  & Opi[n]on-driven and mostly grounded in \textbf{personal experience} -- not viewed from a more general perspective. Quite often focused on particular business goals\dots [GC=1]\\
        \hline\noalign{\smallskip}
    \end{tabular}
\end{table}

\subsection{Professional experience}

Table \ref{table:comments-relating-to-bias-and-honesty}c) presents comments on professional experience.  As noted in the previous subsection, experience is one type of content in blog posts. Table \ref{table:comments-relating-to-bias-and-honesty}c) suggests that experience seems to be the type of content typically of most interest to researchers.

\subsection{The credibility of the author}

Table \ref{table:comments-relating-to-author-credibility} presents comments on the credibility of the author. Respondents appear to place significant value on the credibility of the author of the blog post. The table provides indications of a complex set of attributes that determine credibility i.e., the more experienced and skilled practitioners, who reflect carefully on their work, who are appropriately motivated, who have a high/er reputation, who write regularly on a topic, who are affiliated with a well--known organisation, and who have some history of academic participation.

\begin{table}[!htpb]
    \small
    \center
    \caption{The credibility of the author (grouped by GC score)}
    \label{table:comments-relating-to-author-credibility}
    \begin{tabular}{ p{0.3cm}  p{14.2cm} } 
        \hline\noalign{\smallskip}
        \textbf{\texttt{Q\#}} & \textbf{Comment}\\
        \hline\noalign{\smallskip}

        % GC = 4

        \refstepcounter{rowcount} \therowcount \label{quote:empty17} & I tend to judge articles on the \textbf{credibility of the authors}.  In this case, the \textbf{best practitioners} are absolutely credible.  However, no one is right all the time. [GC=4]\\

        \refstepcounter{rowcount} \therowcount \label{quote:empty18}  & I also appreciate if the \textbf{prior beliefs/assumptions of the author(s)} are made clear in the article. [GC=4]\\

        \refstepcounter{rowcount} \therowcount \label{quote:empty19} & Implicitly, the \textbf{popularity of  the authors}, the \textbf{popularity of the blog} (is it well known, can everyone post there or was the author accepted to present his content there). [GC=4]\\
        \multicolumn{2}{l}{}\\
        
        % GC = 3

        \refstepcounter{rowcount} \therowcount \label{quote:empty20}  & The writer \textbf{experience} (previous worked companies, for instance) The blog credibility (is it part of a larger portal?) [GC=3]\\
        
        \refstepcounter{rowcount} \therowcount \label{quote:empty21}  & I think we should assess the type of information according to its [the author's] \textbf{intention}. It is unfair to assess a blog post as we would do to a scientific paper, and vice--versa. They have \textbf{different target audience}. I understand that blog posts that \textbf{present some empirical data} or \textbf{method} [of data collection] have additional positive characteristics, while scientific works that present neither of them have additional negative characteristics. Both can be assessed regarding a their \textbf{reasoning} and the use of \textbf{practical experiences}, though. [GC=3]\\

        \refstepcounter{rowcount} \therowcount \label{quote:empty22} & It depends on \textbf{who wrote the blog post}. I'm highly biased by blog posts by \textbf{Martin Folwer} (sic) and others. But I read with care posts by other that I don't know. [GC=3]\\

        \refstepcounter{rowcount} \therowcount \label{quote:empty23} & It really depends on \textbf{who is the practitioner}, and the \textbf{blog itself}. If it is part of a larger portal, or independent. [GC=3]\\

        \refstepcounter{rowcount} \therowcount \label{quote:empty24}  & The \textbf{prestige} of the blogger, e.g. Kent Beck [GC=3]\\

        \refstepcounter{rowcount} \therowcount \label{quote:empty25} & Unlike academic articles, which can be judged on the basis of the \textbf{method} and \textbf{data collected}, the quality of a blog articles also depends on the \textbf{reputation} and \textbf{experience} of the author: everyone can have an opinion (which is what most blog articles express I think), but \textbf{opinions of reputable authors} carry more weight. [GC=3]\\

        \refstepcounter{rowcount} \therowcount \label{quote:empty26}  & Author \textbf{affiliation} and \textbf{expertise}. Martin Fowler vs a less experienced [person] [GC=3]\\

        \refstepcounter{rowcount} \therowcount \label{quote:empty27} & Frank disclosure of the writer's \textbf{background and experience} that has shaped its understanding of the world and software development processes. [GC=3]\\

        \refstepcounter{rowcount} \therowcount \label{quote:empty28}  & If \textbf{a practitioner is writing about something regularly}, then any post they write within that area is much more credible than a post in an unrelated or even just different area. [GC=3]\\

        \refstepcounter{rowcount} \therowcount \label{quote:empty29}  & \textbf{Motivation of the author}, for example: promoting company technology/practices, consultant looking for publicity. [GC=3]\\

        \refstepcounter{rowcount} \therowcount \label{quote:reputation-1} & \textbf{Reputation} of the author. [GC=3]\\
        \multicolumn{2}{l}{}\\
        
        % GC = 2

        \refstepcounter{rowcount} \therowcount \label{quote:empty30} & I tend to trust more blogs by \textbf{practitioners of well known companies}. [GC=2]\\

        \refstepcounter{rowcount} \therowcount \label{quote:reputation-2} & \textbf{Reputation} of the blogger. [GC=2] \\

        \refstepcounter{rowcount} \therowcount \label{quote:empty31}  & The \textbf{prior beliefs of the blog writer} - does the writer offer evidence of challenging/\textbf{reflecting upon their own personal beliefs} in response to any learning/understanding they have acquired on the topic they have written the blog on. [GC=2]\\
        \multicolumn{2}{l}{}\\
        
        % GC = 1
 
        \refstepcounter{rowcount} \therowcount \label{quote:empty32} & Quite hard. \textbf{Only well known sources} [authors] and depending on the \textbf{topic} has some minimal credibility. [GC=1]\\
 
        \refstepcounter{rowcount} \therowcount \label{quote:empty33} & practitioner previous \textbf{academic participation} [and] \textbf{practitioner affiliation}. [GC=1]\\
        
        \hline\noalign{\smallskip}
\end{tabular}
\end{table}

\subsection{Anecdotal, (a)contextual information and evidence}

Table \ref{table:comments-relating-to-context-anecdote-and-evidence} presents comments anecdote and context. The concepts of context and anecdote are related to each other in that anecdotal evidence appears to lack (amongst other things) important details, such as contextual information and concrete detail. Some respondents express concerns that practitioners in blog posts over--generalise beyond their experiences.

Anecdote and context have connections with case studies (e.g. \citep{runeson2012case, runeson2009guidelines}) and their generalisation. One recurring threat to the validity of case studies, and also of survey studies, is the limited number of cases reported. Another threat is that often researchers simply do not know the characteristics of the population from which the cases are drawn. And a further threat is that the sampling process itself is uncertain e.g., self--selection on the part of respondents. These threats, and the recurring problem of context in software engineering research (e.g. \citep{petersen2009context, dybaa2012works, dyba2013contextualizing, clarke2012situational}) raise concerns about generalisation.

One difference between the anecdotes of practitioners and the case study by researchers is that researchers are much more likely to recognise and report these challenges, in contrast to practitioners writing blog posts. Simply reporting the presence of a threat does not in itself diminish the threat, however. Even when researchers report contextual information, the discipline lacks an accepted framework for sharing contextual information.

\begin{table}[!htpb]
    \small
    \center
    \caption{Context and anecdote (grouped by GC score)}
    \label{table:comments-relating-to-context-anecdote-and-evidence}
    \begin{tabular}{ p{0.3cm}  p{14.2cm} } 
    \hline\noalign{\smallskip}
    \textbf{\texttt{Q\#}} & \textbf{Comment}\\
    \hline\noalign{\smallskip}
    
    % GC = 4

    \refstepcounter{rowcount} \therowcount \label{quote:relevant} & \textbf{Anecdotal}, but surely relevant. [GC=4]\\
    \multicolumn{2}{l}{}\\
    
    % GC = 3
    
    \refstepcounter{rowcount} \therowcount \label{quote:empty34} & Many practitioners lack \textbf{the larger picture} to give objective reports. As such, they report \textbf{their perspective} which \textbf{is often skewed} by their limited knowledge of tools and techniques\dots As such, vast majority of practitioners are behind the state of art and practice and report \textbf{anecdotal} or otherwise \textbf{out--of--context} findings that should not be considered by anyone thriving towards excellence in any field. On the other hand, there are a select few practitioners who have sufficient background knowledge and larger perspective whose blogs are excellent and far beyond any researcher's blogs or in some cases even surpassing scientific journals in their value. These practitioners always: 1) \textbf{avoid generalisation}, and 2) \textbf{report in detail} sufficient for reproduction, verification and scoping (as they do list the \textbf{biases} in their analysis, which rarely happens in journal articles). [GC=3]\\

    \refstepcounter{rowcount} \therowcount \label{quote:empty35} & Blog entries (e.g. position statements / manifestos) are to be seen as \textbf{anecdotal evidence}. This is not a problem per--se, but such evidence suffers especially from the \textbf{lack of context} information that makes the claims empirically usable rather than suffering from the lack of evidence-based accuracy [GC=3]\\

    \refstepcounter{rowcount} \therowcount \label{quote:empty36} & \textbf{Context} description (good description) of the blog, e.g. Agile text can be many versions and variants of Agile. [GC=3]\\
    
    \refstepcounter{rowcount} \therowcount \label{quote:empty37} & The \textbf{context information} (e.g. project setting, practitioners' characteristics, etc.) is rather important. Setting the results [practitioner's observations and beliefs] in relation to \textbf{existing evidence} can be done by researchers provided [there is] a certain clarity of the [practitioner's] claims. Without clarity, claims tend to result in universal propositions and solutions. [GC=3]\\
    \multicolumn{2}{l}{}\\
    
    % GC = 2

    \refstepcounter{rowcount} \therowcount \label{quote:empty38} & Blogs contain personal opinions that stem from \textbf{concrete cases}. They cannot be generalized. Furthermore, opinions evolve over time as technology matures, the writer acquires more experience, etc. In other words: blogs are snapshots (on time / \textbf{contex[t]}) regarding some issue / technology. [GC=2]\\

    \refstepcounter{rowcount} \therowcount \label{quote:empty39} & Some blogs are good (e.g., Jeff Atwood or Martin Fowler). However, most of those are not scientifically validated. Does not contain enough information to understand the \textbf{context} or replicate. [GC=2]\\
    \multicolumn{2}{l}{}\\

    % GC = 1

    \refstepcounter{rowcount} \therowcount \label{quote:empty40} & Very few are credible. Main problem is they are \textbf{most often anecdotal} and based on \textbf{often a limited set of cases} where the blog authors has worked/\textbf{experience} from. Since this is \textbf{their experience} there is rarely a realization that it is limited. I know practitioners who spent their whole adult work life in one and the same company and this gives them the right to know much more and ``correct'' researcher even if the latter are working closely with maybe 10--15 different companies. I think \textbf{we often think way too binary on these things} (``all practitioners know a lot about industrial practice, while no researchers do''). [GC=1l; parenthesis are in the original text]\\

    \refstepcounter{rowcount} \therowcount \label{quote:empty41} & Reporting the \textbf{context} for and \textbf{how empirical evidence has been gathered} is the key thing. \textbf{Anecdotal ``stories''} based on long \textbf{experience} can be helpful to create ideas of what to study more but can and often is misused by authors to spread their opinions. [GC=1]\\

    \refstepcounter{rowcount} \therowcount \label{quote:empty42} & some blogs are written by practitioners who have \textbf{access to data} from the company they work for (see StackOverflow blog). That is very valuable knowledge. Many others (especially the ones not explicitly affiliated) are less reliable since based on \textbf{anecdotes}. [GC=1]\\

    \refstepcounter{rowcount} \therowcount \label{quote:empty43} & Describing the \textbf{context} for ones statements/blog/info. What is your experience, how broad is it, from which industries/domains, how many years etc. See Peterseon and Wohlin on describing context for empirical SE research, for example. [GC=1]\\
    
    \hline\noalign{\smallskip}
\end{tabular}
\end{table}

Table \ref{table:comments-relating-to-context-anecdote-and-evidence} presents verbatim quotes on the relationship between anecdote and empirical evidence. Essentially, the respondent is suggesting we seek alignment -- a triangulation -- between the beliefs reported by the practitioner, one's professional experience (as a practitioner) and independent empirical evidence. This is the intended aim of Multivocal Literature Reviews (MLRs) and also of Evidence Based Software Engineering (EBSE). 

\section{Preliminary checklists for evaluating blog posts}
\label{section:checklist}

\subsection{Overview}

Building on the analyses presented earlier in this paper, we propose a preliminary checklist that researchers can use to evaluate blog posts. We also briefly demonstrate the checklist in section \ref{section:comparison}.

The checklist is synthesised from four sources: the general review of literature we present in section \ref{section:related-work}, the pre--defined criteria used in the survey (see section \ref{subsection:candidate-criteria} and also \citep{williams_rainer_cred_tr}), the thematic analyses of the reviewers' qualitative responses (see section \ref{section:qualitative-results}), and two quality checklists, Garousi \textit{et al}.'s \citeyearpar{garousi2018guidelines} generic checklist for multivocal literature reviews (MLRs), and Soldani \textit{et al}.'s \citeyearpar{soldani2018pains} checklist used in their grey literature review (GLR) of microservices. The checklist is organised into sections as summarised in Table \ref{table:checklist-categories}. The checklist itself is presented in Tables \ref{table:foundation-checklist} through \ref{table:checklist-continued-4}.

\begin{table}[!htpb]
    \small
    \center
    \caption{Summary of the checklists}
    \label{table:checklist-categories}
    \begin{tabular}{ | p{1.8cm} | p{13.2cm} | }
    \hline
    \textbf{Checklists} & \textbf{Brief explanation}\\
    \hline
    Foundational & Questions intended to raise the self--awareness of users of the other checklists.\\
    Process & Questions for evaluating the process of enquiry -- if any -- that informs the writing of the blog posts, e.g., any formal or informal investigation conducted by the author of the blog post\\
    Author & Questions for evaluating the credibility of the author of the blog post.\\
    Content & Questions for evaluating the content of the blog post itself.\\
    Reader & Questions for evaluating reader feedback in the blog post.\\
    Influencer & Questions for evaluating influences of others on the author of the blog post.\\
    Media & Questions for evaluating the media through which the blog post is published, e.g., the platform on which the blog post is posted.\\
    \hline
    \end{tabular}
\end{table} 

All of the sections of the checklist are concerned with credibility (broadly defined) and relevance, though some sections are more explicit in their separation of these two constructs, and some sections focus on a particular aspect of credibility. Some sections are also more developed the others, and we briefly discuss three sections here: the Foundational, Process and Author sections.

For the Foundational section, the worldview provides a foundation for framing the topic, objectives, and questions for the enquiry, the features of the empirical phenomena of interest, and the foundations of how credibility will be evaluated. Individuals have worldviews. A worldview may also be implicitly or explicitly stated in guidelines, checklists or methodology.

The Process section is a special case because it focuses only on the rigour on an enquiry process and not on the relevance.

For the Author section, credibility is demonstrated and assessed through competence and relevance e.g., that the author has sufficient expertise in a relevant domain. Researchers select practitioners for studies, or allow practitioners to self--select, on the basis of competence and relevance. Researchers have concerns around the subjectivity and hence bias of the enquirer, and this raises a paradox: researchers appear to value (some) practitioners on the basis of their professional experiences and the reporting of that experience (e.g., Martin Fowler) but also have concerns around the subjectivity that arises during the very formation of such experiences and their reporting.

We deliberately do not provide a scoring mechanism or thresholds, or similar, for the checklist. Neither do we advise on how to aggregate `scores' within the checklist. We do this in recognition of the insights arising from our qualitative and quantitative analyses, and in recognition of Petersen and Gencel's (\citeyear{petersen2013worldviews} discussion of the relationship of worldviews and validity, e.g., to set a scoring mechanism would be to prematurely set a standard for assessing a complex type of evidence. 

We present a preliminary checklist because the checklist would benefit from further development, validation and evaluation. The checklist therefore constitutes a foundation for further research, e.g., through its application, evaluation and revision.

\begin{table}[!htpb]
    \small
    \center
    \caption{Checklist}
    \label{table:foundation-checklist}
    \begin{tabular}{p{0.3cm} p{14.6cm}} 
        % \hline\noalign{\smallskip}
        \textbf{\#} & \textbf{Question}\\
        \hline\noalign{\smallskip}
        
        \multicolumn{2}{l}{\textbf{Foundational}}\\
        % & \textbf{Foundational} \\
        
        \multicolumn{2}{l}{Familiarity and expectations of blog posts}\\
        % & Familiarity and expectations of blog posts\\
        
        \refstepcounter{checklistItem} \arabic{checklistItem} \label{c1} & How frequently do you read blog posts?\\
        
        \refstepcounter{checklistItem} \arabic{checklistItem} \label{c2} & Have you used blog posts in your research? \\           
        \refstepcounter{checklistItem} \arabic{checklistItem} \label{c3} & What are your prior beliefs about blog posts? \\
         
        \multicolumn{2}{l}{Worldview}\\
         
        % & Worldview\\
        
        \refstepcounter{checklistItem} \arabic{checklistItem} \label{c4} & What is your worldview of research, e.g., pragmatic, constructivist, positivist? \\
         
        \refstepcounter{checklistItem} \arabic{checklistItem} \label{c5} & Have you reflected on your worldview? \\
         
        \refstepcounter{checklistItem} \arabic{checklistItem} \label{c6} & Have you reflected on how your worldview influences, or might influence, your approach to the evaluation of blog posts? \\

        \multicolumn{2}{l}{Research design}\\

        % & Research design\\

        \refstepcounter{checklistItem} \arabic{checklistItem} \label{c7} & These checklists complement other checklists. \\
         
        \refstepcounter{checklistItem} \arabic{checklistItem} \label{c8} & Have you clearly defined the empirical phenomenon, or phenomena, that interests you and to which the blog posts are expected to relate? \\
         
        \refstepcounter{checklistItem} \arabic{checklistItem} \label{c9} & Have you clearly defined the \textit{context} for the empirical phenomenon, or phenomena, that interests you and to which the blog posts are expected to relate?? \\
         
        \refstepcounter{checklistItem} \arabic{checklistItem} \label{c10} & Have you a clear conceptual model of the theorised conditions that give rise to the empirical phenomena that interest you? \\
         
        \refstepcounter{checklistItem} \arabic{checklistItem} \label{c11} & Have you developed a clear set of research questions, or similar, to direct your evaluation? \\
        
        % \multicolumn{2}{l}{}\\
        
    \end{tabular}
\end{table}  

\begin{table}[!htpb]
    \small
    \center
    \caption{Checklist (continued)}
    \label{table:checklist-continued-2}
    \begin{tabular}{p{0.4cm} p{14.5cm}} 

        \textbf{\#} & \textbf{Question}\\
        \hline\noalign{\smallskip}

\multicolumn{2}{l}{\textbf{Process}}\\

        %  & \textbf{Process} \\
        
        \refstepcounter{checklistItem} \arabic{checklistItem} \label{c12} & Is there a process of enquiry reported in the blog post at least in some general way?\\        
        \refstepcounter{checklistItem} \arabic{checklistItem} \label{c13} & Are specific steps of the enquiry reported, in particular is there some explanation for:\\
        \arabic{checklistItem}.1 & \quad how observations were made;\\
        \arabic{checklistItem}.2 & \quad how data was collected;\\
        
        \arabic{checklistItem}.3 & \quad how data was analysed;\\
        
        \arabic{checklistItem}.4 & \quad how results were interpreted; and\\
        
        \arabic{checklistItem}.5 & \quad how interpretations were reported? \\
 
        \multicolumn{2}{l}{}\\
        
        \refstepcounter{checklistItem} \arabic{checklistItem} \label{c14} & Are the steps of the process reported accurately?\\
        
        \refstepcounter{checklistItem} \arabic{checklistItem} \label{c15} & Are the steps of the process reported completely?\\
        
        \refstepcounter{checklistItem} \arabic{checklistItem} \label{c16} & Does the process conform to available standards?\\
        
        \refstepcounter{checklistItem} \arabic{checklistItem} \label{c17} & Is the process fit--for--purpose?\\
        
        \refstepcounter{checklistItem} \arabic{checklistItem} \label{c18} & Is the process reported completely?\\
        
        \refstepcounter{checklistItem} \arabic{checklistItem} \label{c19} & Is the process conducted in a timely way?\\
        
        \refstepcounter{checklistItem} \arabic{checklistItem} \label{c20} & Is the process reported in a timely way?\\

        % \multicolumn{2}{l}{}\\
        
        \refstepcounter{checklistItem} \arabic{checklistItem} \label{c21} & Does the enquiry process support realism?\\
        
        \refstepcounter{checklistItem} \arabic{checklistItem} \label{c22} & Does the enquiry process use realistic subjects? \\
        
        \refstepcounter{checklistItem} \arabic{checklistItem} \label{c23} & Does the enquiry process occur in a realistic context?\\
        
        \refstepcounter{checklistItem} \arabic{checklistItem} \label{c24} & Does the enquiry process investigate a phenomena at an appropriate scale?\\
        
        % \multicolumn{2}{l}{}\\
        % \multicolumn{2}{l}{\textbf{Author}}\\

        % %  & \textbf{Author} \\
        % \multicolumn{2}{l}{Reputation:}\\
        % % & Reputation:\\
        % \refstepcounter{checklistItem} \arabic{checklistItem} \label{c25} & What is the author's professional reputation?\\
        % \refstepcounter{checklistItem} \arabic{checklistItem} \label{c26} & What is the author's affiliation?\\
        % \refstepcounter{checklistItem} \arabic{checklistItem} \label{c27} & What is the affiliation's reputation?\\

        % \refstepcounter{checklistItem} \arabic{checklistItem} \label{c28} & How regularly does the author publish posts?\\
 
        % \multicolumn{2}{l}{Bias, subjectivity, intentions \& sincerity:}\\
        
        % % & Bias, subjectivity, intentions \& sincerity:\\

        % \refstepcounter{checklistItem} \arabic{checklistItem} \label{c29} & What are the intentions of the author?\\
        
        % \refstepcounter{checklistItem} \arabic{checklistItem} \label{c30} & Does the author have any bias?\\
        
        % \refstepcounter{checklistItem} \arabic{checklistItem} \label{c31} & Does the author have vested interests?\\
        
        % \refstepcounter{checklistItem} \arabic{checklistItem} \label{c32} & Is the author sincere?\\
        
        % \refstepcounter{checklistItem} \arabic{checklistItem} \label{c33} & Is there any relevant subjectivity?\\      

        % \multicolumn{2}{l}{Expertise and experience}\\
        
        % % & Expertise and experience\\

        % \refstepcounter{checklistItem} \arabic{checklistItem} \label{c34} & What is the author's relevant experience?\\
        
        % \refstepcounter{checklistItem} \arabic{checklistItem} \label{c35} & What is the author's relevant expertise? \\

        \hline\noalign{\smallskip}

        \hline\noalign{\smallskip}
    \end{tabular}
\end{table}  

\begin{table}[!htpb]
    \small
    \center
    \caption{Checklist (continued)}
    \label{table:checklist-continued-3}
    \begin{tabular}{p{0.4cm} p{14.5cm}} 

        \textbf{\#} & \textbf{Question}\\
        \hline\noalign{\smallskip}

\multicolumn{2}{l}{\textbf{Author}}\\

        %  & \textbf{Author} \\
        \multicolumn{2}{l}{Reputation:}\\
        % & Reputation:\\
        \refstepcounter{checklistItem} \arabic{checklistItem} \label{c25} & What is the author's professional reputation?\\
        \refstepcounter{checklistItem} \arabic{checklistItem} \label{c26} & What is the author's affiliation?\\
        \refstepcounter{checklistItem} \arabic{checklistItem} \label{c27} & What is the affiliation's reputation?\\

        \refstepcounter{checklistItem} \arabic{checklistItem} \label{c28} & How regularly does the author publish posts?\\
 
        \multicolumn{2}{l}{Bias, subjectivity, intentions \& sincerity:}\\
        
        % & Bias, subjectivity, intentions \& sincerity:\\

        \refstepcounter{checklistItem} \arabic{checklistItem} \label{c29} & What are the intentions of the author?\\
        
        \refstepcounter{checklistItem} \arabic{checklistItem} \label{c30} & Does the author have any bias?\\
        
        \refstepcounter{checklistItem} \arabic{checklistItem} \label{c31} & Does the author have vested interests?\\
        
        \refstepcounter{checklistItem} \arabic{checklistItem} \label{c32} & Is the author sincere?\\
        
        \refstepcounter{checklistItem} \arabic{checklistItem} \label{c33} & Is there any relevant subjectivity?\\      

        \multicolumn{2}{l}{Expertise and experience}\\
        
        % & Expertise and experience\\

        \refstepcounter{checklistItem} \arabic{checklistItem} \label{c34} & What is the author's relevant experience?\\
        
        \refstepcounter{checklistItem} \arabic{checklistItem} \label{c35} & What is the author's relevant expertise? \\

        \hline\noalign{\smallskip}
    \end{tabular}
\end{table}  

\begin{table}[!htpb]
    \small
    \center
    \caption{Checklist (continued)}
    \label{table:checklist-continued-4}
    \begin{tabular}{p{0.4cm} p{14.5cm}} 

        \textbf{\#} & \textbf{Question}\\
        \hline\noalign{\smallskip}
        \multicolumn{2}{l}{\textbf{Content}}\\
        %  & \textbf{Content} \\
        \multicolumn{2}{l}{Quality of writing }\\
        % & Quality of writing \\

        \refstepcounter{checklistItem} \arabic{checklistItem} \label{c36} & How clearly written is the post? \\

        \refstepcounter{checklistItem} \arabic{checklistItem} \label{c37} &How impartially written is the post? \\

        \multicolumn{2}{l}{Anecdotal content}\\

        % & Anecdotal content \\

        \refstepcounter{checklistItem} \arabic{checklistItem} \label{c38} & Does the post present anecdote? \\
        \refstepcounter{checklistItem} \arabic{checklistItem} \label{c39} & Is the anecdote supported by professional experience?\\
        \refstepcounter{checklistItem} \arabic{checklistItem} \label{c40} & Does the post report concrete details of the anecdote? \\
        \refstepcounter{checklistItem} \arabic{checklistItem} \label{c41} & Are concrete details reported about discrete situations?\\
        
        \refstepcounter{checklistItem} \arabic{checklistItem} \label{c42} & Does the post report context information?\\

        \multicolumn{2}{l}{Empirical content}\\
        
        % & Empirical content \\
        
        \refstepcounter{checklistItem} \arabic{checklistItem} \label{c43} & Does the post report empirical data?\\   
        
        \multicolumn{2}{l}{Practical content}\\

        % & Practical content\\

        \refstepcounter{checklistItem} \arabic{checklistItem} \label{c44} & Does the post report a Minimum, Complete and Verifiable Example (MCVE)?\\
        
        \refstepcounter{checklistItem} \arabic{checklistItem} \label{c45} & Does the post report source code example(s)?\\

        \multicolumn{2}{l}{Conceptual content}\\

        % & Conceptual content\\

        \refstepcounter{checklistItem} \arabic{checklistItem} \label{c46} & Does the post report a conceptual model or similar? \\

        \multicolumn{2}{l}{Rhetorical content}\\

        % & Rhetorical content\\

        \refstepcounter{checklistItem} \arabic{checklistItem} \label{c47} & Does the post make claims about relevant professional situations?\\
        
        \refstepcounter{checklistItem} \arabic{checklistItem} \label{c48} &  Are claims supported by reasons, explanations etc?\\

        \refstepcounter{checklistItem} \arabic{checklistItem} \label{c49} &  Are claims supported by anecdotal content?\\
 
        \refstepcounter{checklistItem} \arabic{checklistItem} \label{c50} &  Are claims supported by empirical content?\\

        \refstepcounter{checklistItem} \arabic{checklistItem} \label{c51} &  Are claims supported by practical content?\\

        \refstepcounter{checklistItem} \arabic{checklistItem} \label{c52} &  Are claims supported by conceptual content?\\

        \refstepcounter{checklistItem} \arabic{checklistItem} \label{c53} &  Does the post show a clear connection between claims, rhetoric, conceptual, and/or anecdotal or empirical content \\
        
        \multicolumn{2}{l}{\textbf{Reader}}\\

        %  & \textbf{Reader} \\
        \refstepcounter{checklistItem} \arabic{checklistItem} \label{c54} & How many times has the post been cited by others (backlink)?\\
        %\hline
        % & Number of backlinks & & & & GQ & Yes &\\
        %\hline
        \refstepcounter{checklistItem} \arabic{checklistItem} \label{c55} & How many times has the post been media--shared\\
        %\hline
        \refstepcounter{checklistItem} \arabic{checklistItem} \label{c56} & How many comments does the post have?\\
        %\hline
        \refstepcounter{checklistItem} \arabic{checklistItem} \label{c57} & How many times has the post been viewed?\\
        
        \multicolumn{2}{l}{\textbf{Influencer}}\\
        \multicolumn{2}{l}{Influence of practitioners' written work}\\

        %  & \textbf{Influencer} \\
        
        % & Influence of practitioners' written work\\
     
        \refstepcounter{checklistItem} \arabic{checklistItem} \label{c58} & How many citations (e.g., backlinks) are there to practitioner's published work\\

        \multicolumn{2}{l}{Influence of researchers' written work}\\

        % & Influence of researchers' written work\\
        
        \refstepcounter{checklistItem} \arabic{checklistItem} \label{c59} & How many citations (e.g., references) are there to researcher's published work\\
        
         \multicolumn{2}{l}{\textbf{Media}}\\

        %  & \textbf{Media} \\

        \refstepcounter{checklistItem} \arabic{checklistItem} \label{c60} & Medium \\
        \refstepcounter{checklistItem} \arabic{checklistItem} \label{c61} & Platform \\
        \hline\noalign{\smallskip}
    \end{tabular}
\end{table}

\subsection{Applying the checklist to the survey itself}
\label{section:comparison}

% To gain further insight into the credibility of blog posts, we use the models from section \ref{section:models-and-evaluation} to explore characteristics of the responses in our survey. We consider both the content of particular survey responses and also the more general tendencies in the responses. This exploration also partially demonstrates the application of the models.

Although the checklists were developed to evaluate blog posts, we demonstrate the checklists by applying them to the survey responses. We apply the checklists to the survey for several reasons. First, it is more efficient to demonstrate, in this paper, the checklists against the dataset we have already been exploring, rather than introduce and discuss a further dataset. Second, applying the checklists to the survey is another way of evaluating the survey data. Third, by applying the checklists to the survey data, we begin to show that the checklists have the potential for wider applicability. Fourth, and finally, by applying the checklists to the survey data we can contrast the credibility of survey data with the credibility of blog posts as data.

There is the risk in this demonstration of a circular argument: we use the survey data to generate checklists and then use the checklists to evaluate the same survey data. To clarify: we have used the content of the survey responses to generate the checklists, and now use the checklists to examine potential contradictions in the survey data and to compare that data with blog posts.

\begin{table}[!htpb]
\centering
\small
\caption{Characteristics of the responses from survey respondents}
\label{table:exploration-of-responses}
\begin{tabular}{ p{3cm}  p{12cm} }
\hline\noalign{\smallskip}
\textbf{Criteria} & \textbf{Examples}\\
\hline\noalign{\smallskip}
\multicolumn{2}{l}{\textit{Items from the Foundational category}}\\
\hline\noalign{\smallskip}
Worldview & Table \ref{table:comparison-with-interviews} suggests that some researchers conceive of blog posts as `types' of interview.\\
Values & Some responses suggest that respondents are trading--off the value of rigour and relevance e.g., \texttt{Q\ref{quote:relevant}}.\\
\hline\noalign{\smallskip}
\multicolumn{2}{l}{\textit{Items from the Process category}}\\
\hline\noalign{\smallskip}
Report & We show in Table \ref{table:statistics-on-researchers-responses} that the `reports' from respondents are short, even very short.  Table \ref{table:statistics-on-Joel-Spolsky} presents contrasting statistics for the set of blog posts on the Joel Spolsky blog.\\
Steps of the enquiry & Responses such as those presented in Table \ref{table:general-observations} imply that some respondents have not conducted \textit{any} kind of enquiry into blog posts, however informal.\\
\hline\noalign{\smallskip}
\multicolumn{2}{l}{\textit{Items from the Author category}}\\
\hline\noalign{\smallskip}
Experience & Table \ref{table:general-observations} provides examples of where respondents declare they lack experience. And yet, as far as we are aware,\textit{no} respondent explicitly considered themselves unqualified to complete the survey. In the follow--up with non--respondents to better understand why they did not participate in (did not start, or did not complete) the survey, the \textit{only} declared reason for not participating in the survey was the lack of time.\\
\hline\noalign{\smallskip}
Vested interests & By definition, the respondents have vested interests as they are being asked to evaluate the credibility of blog posts for use in research in their discipline.\\
\hline\noalign{\smallskip}
\multicolumn{2}{l}{\textit{Items from the Content category}}\\
\hline\noalign{\smallskip}
Concreteness & Respondents referred to particular practitioners e.g., in \texttt{Q\ref{quote:don't-follow-blogs}} the respondent refers to, ``\dots\space the work of practitioners such as Don Reifer, Capers Jones, Larry Putnam, Dan Galorath; organizations such as NESMA and IFPUG; and data providers such as ISBSG.''\\
\noalign{\smallskip}
Reasoning & There are short responses (e.g. \texttt{Q\ref{quote:quality-varies}}, \texttt{Q\ref{quote:reputation-1}}, \texttt{Q\ref{quote:reputation-2}}) that contain assertions that are not supported by explanations, arguments, citations etc.\\
\noalign{\smallskip}
Experience & Some respondents refer to their personal experience, but don't describe that experience e.g., \texttt{Q\ref{quote:my-experience-1}, Q\ref{quote:my-experience-2}}.\\
\noalign{\smallskip}
Beliefs & Some respondents refer to their beliefs e.g., \texttt{Q\ref{quote:believe-1}, Q\ref{quote:believe-2}, Q\ref{quote:believe-3}}.\\
\noalign{\smallskip}
Impartiality of writing & In general the responses are (relatively) impartial, though there are some exceptions e.g., \texttt{Q\ref{quote:impartiality-1}}.\\
\noalign{\smallskip}
Influence of others & Considering that all 44 respondents were researchers, only \textit{one} response refers to \textit{research} (and even then with no explicit citation): ``\dots\space See Peterseon (sic) and Wohlin on describing context for empirical SE research\dots''. (We presume the respondent was referring to \citep{petersen2009context}.)\\
\noalign{\smallskip}
& A greater number of responses explicitly recognised specific practitioners, the most frequent being Martin Fowler. Again, there were no explicit citations e.g. no URLs were provided.  \\
\hline\noalign{\smallskip}
\end{tabular}
\end{table}

\begin{table}[ht]
    \small
    \center
    \caption{Statistics on researchers' responses}
    \label{table:statistics-on-researchers-responses}
    \begin{tabular}{ l  r } 
        \hline\noalign{\smallskip}
        \textbf{Statistic} & \textbf{Value}\\
        \hline\noalign{\smallskip}
        %\hline
        Total number of words (across all questions) & 3,972 \\
        Total actual number of qualitative responses received (across all questions) & 143\\
        Maximum number of potential qualitative responses  (across all questions) & 264\\
        Percentage of actual number of qualitative responses: & 54\%\\
        Mean number of words per response & 15\\
        Median number of words per response & 5\\
        Total number of reasoning markers & 33\\
        Mean number of reasoning markers (264 responses) & 0.125\\
        Median number of reasoning markers & 0\\
        \hline\noalign{\smallskip}
    \end{tabular}
\end{table}

\begin{table}[ht]
    \small
    \center
    \caption{Statistics on Joel Spolsky's blog posts (n=1024)}
    \label{table:statistics-on-Joel-Spolsky}
    \begin{tabular}{ l  r  r } 
    \hline\noalign{\smallskip}
    \textbf{Statistic} & \textbf{Excluding outlier} & \textbf{Including outlier}\\
    \hline\noalign{\smallskip}
    Minimum number of words & 0 & 0 \\
    Maximum number of words & 6222 & 16432 \\
    Mean number of words & 515.3 & 530.7\\
    Median number of words& 181 & 181.5\\
    \hline\noalign{\smallskip}
    \multicolumn{3}{l}{Note: The minimum count of zero words is due to the post containing a photo.} 
    \end{tabular}
\end{table}

\begin{table}[!htpb]
    %\centering
    \small
    \center
    \caption{Blog posts compared with interviews}
    \label{table:comparison-with-interviews}
    \begin{tabular}{ p{0.5cm}  p{14.5cm} } 
        \hline\noalign{\smallskip}
        \texttt{\textbf{Q\#}} & \textbf{Researchers' expectations}\\
        \hline\noalign{\smallskip}
        %\hline
        \refstepcounter{rowcount} \therowcount \label{quote:emptycomparison1} & Blog articles are credible as a witness of a practitioner, and has to be treated accordingly. I would never publish a practitioner interview in extenso, neither is a practitioner blog post to be trusted as rigorous research outputs. [GC=2]\\
        %\hline
        \refstepcounter{rowcount} \therowcount \label{quote:emptycomparison2}& I consider them mostly equivalent to interviews. While practitioners are very free to edit their words, they still represent their opinions/perceptions/emotions. [GC=4]\\
        \hline\noalign{\smallskip}
    \end{tabular}
\end{table}

Table \ref{table:exploration-of-responses} summarises criteria from the models with illustrative examples from the survey responses. Three further tables support Table \ref{table:exploration-of-responses}: Table \ref{table:statistics-on-researchers-responses} presents statistics on the responses from survey respondents; Table \ref{table:statistics-on-Joel-Spolsky} presents statistics on blog posts from one blog maintained by Joel Spolsky; and Table \ref{table:comparison-with-interviews} provides quotes from the survey respondents on the nature of blog posts. For Table \ref{table:statistics-on-researchers-responses}, the reasoning markers are discussed and validated in \citep{williams2018using}. For Table \ref{table:statistics-on-Joel-Spolsky}, we have used the Joel Spolsky dataset in previous work e.g., \citep{williamsEASE19practitionerscite,williams2018software}.

The comparisons presented in Table \ref{table:exploration-of-responses} through Table \ref{table:comparison-with-interviews} suggest that respondents to the survey are behaving a bit like practitioners who write blog posts. Respondents are reporting their beliefs, but these beliefs are often supported by limited \textit{reporting of} experience, reasoning, anecdote etc.; or are even unsupported. We have found instances where respondents self--report a lack of experience of blog posts, and yet continue to provide opinions on the credibility of blog posts: these respondents should probably be excluded from the analyses. As another example, many survey respondents value reasoning, and yet there are many statements made in the survey that are not supported by reasoning but, instead, are simply assertions. Despite these constraints and limitations in the data, we have still been able to develop a meta--model and several simple models. A similar approach applies to blog posts: though individual blog posts have constraints and limitations on quality, it remains possible to aggregate insights from an appropriate set of blog posts, as Soldani \textit{et al}. \citeyearpar{soldani2018pains} and Raulamo--Jurvanen \textit{et al}. \citeyearpar{RaulamoJurvanen2017choosing} have done.

\section{Discussion}
\label{section:discussion}

% \subsection{\textcolor{blue}{Items to add}}
% Items to add include:
% \begin{enumerate}
%     \item Discussion of GLRs, SRs and case--survey
% \end{enumerate}

\subsection{Reviewing the research questions}

For question RQ1, \textit{viz.} To what degree do researchers consider blog posts to be credible?, we report a heatmap analysis of the survey responses. Compared to our preliminary analysis \citep{williamsEASE19credibilitysurvey}, this heatmap more concisely and effectively represents the diversity of attitudes within the research community. The heatmap suggests the presence of different worldviews \citep{petersen2013worldviews}, as discussed in section \ref{section:related-work}, and begins to explain the contention present within the research community e.g., between a researcher who values reasoning and relevance in contrast to a researcher who values empirical data and reporting of data collection processes.

For question RQ2, \textit{viz.} What criteria do researchers claim to use when evaluating the credibility of a blog post?, we report the thematic analyses of over 60 qualitative comments contained in the survey (see section \ref{section:qualitative-results}). The comments reveal a complex set of criteria that researchers claim to use when evaluating blog posts. Of course, any individual researcher is unlikely to be using all of these criteria.

For question RQ3, \textit{viz.} What guidance on the quality--assurance of blog posts can be synthesised from prior research and the survey responses?, we have drawn on four sources to synthesis a preliminary suite of checklists: our nine original credibility criteria, the literature review, two quality checklists for grey literature \citep{garousi2018guidelines,soldani2018pains}, and the quantitative and qualitative re--analyses of the survey responses. Each checklist is intended to evaluate a different facet of blog posts.

For question RQ4, \textit{viz.} How do the criteria that the respondents \textit{say} they use compare with the responding \textit{behaviour} of those respondents in the survey itself?, we perform two comparisons. First, we compare characteristics of the survey responses against a dataset of blog posts. Second, we use the synthesised checklists to analyse selected survey responses. We observe that survey respondents' behaviour appears similar in some aspects to the behaviour of authors of blog posts, e.g., survey respondents made assertions and these assertions were rarely supported by argument or reasoning, and researchers did not formally cite previous research to support their assertions. The number of criteria within the checklists indicate the complexity of evaluating the credibility of blog posts, and also partially explains the contrasts in attitudes of the researchers who responded to our survey. Again, the variety of responses from the researchers suggests differences in worldviews, and therefore differences in researchers' research questions and topics of interest, and in the `logic' that researchers use to define and evaluate the credibility of data.

\subsection{Implications of the findings}

The current paper, and the associated survey, were motivated by two issues: the recognition that the use of blog posts as evidence in software engineering is contentious for some, and the recognition of the significant challenge of quality--assuring blog posts. These two issues are clearly connected.

With regards to the first motivation, concerning contention, the findings reported in this paper are not intended to resolve the contention(s) present in the research community; rather, the paper seeks to increase the community's awareness of these issues, as well as the complexity of these issues, and to encourage the community to engage with these issues.

With regards to the second motivation, concerning quality--assurance, the paper develops a preliminary suite of checklists specifically for blog posts to help the community quality--assure blog posts. 

% One contributing factor to this contention is the evidential status of blog posts and the nature of Systematic Reviews.

% Many of the respondents' reservations about the credibility of software practitioners' blog posts may also be directed at other types of content generated by practitioners e.g., content generated through interviews, surveys and focus groups. Some research methods, such as interviews and focus groups, allow for more control and more interaction with the practitioner. This increased control and interaction supports the generation and collection of more rigorous and more relevant information. The checklist developed in section \ref{section:models-and-evaluation} might be used to evaluate the credibility of information collected through these other research methods. 

\subsection{Threats to validity}
\label{subsection:threats}

There are a number of threats to the validity of this study, these threats also indicating directions for future research. We enumerate the responses:
\begin{enumerate}
    \item Whilst the response rate is reasonable (32\%) the absolute number of responses (44) is relatively low. We have managed this threat through retaining all responses and conducting and reporting a rich sample of qualitative responses (over 60 quotes) to complement the quantitative analyses.
    
    \item Kitchenham and Pfleeger (\citeyear{kitchenham2002surveypart5}) advise that survey designers should specify their population of interest. We have implicitly defined a population by targeting members of programme committees for two very well--established conferences in the field. The implied population was therefore empirical software engineering researchers. We could have been both more explicit on that population but also more specific, e.g., the population of empirical software engineering researchers who have some familiarity with blogs and blog posts. 
    
    \item We didn't explicitly ask respondents about their experience with blog posts e.g., the degree to which they read blog posts and follow different blogs. We also didn't explicitly ask respondents about their expectations of blog posts. These limitation prevent us from distinguishing the more informed and experienced respondents from those who are less informed and less experienced. We have considered the exclusion of cases as part of our survey design in section \ref{section:survey-design} and then in our thematic analyses in section \ref{section:qualitative-results}. With this survey being the first survey of its kind, and given the limited sample size, we decided to retain all responses for the current paper, to maximise the insights on this topic. (In our original paper, we excluded one respondents.)
    
    \item A related concern is that we didn't provide guidelines to potential respondents on whether they should exclude themselves, or whether and how they should declare their degree of confidence in their responses. The respondent (see quote \ref{quote:don't-follow-blogs} in Table \ref{table:general-observations}) who stated that she or he does not follow blogs is an example of a respondent who likely should not participate in a future study of this kind. We have managed this threat in several ways: 1) retaining the entire dataset for completeness, 2) reporting a rich sample of quotes, 3) reporting the respondent's overall rating for the credibility of blog posts with each quote, and 4) synthesising a suite of checklists from the quotes.
    
    \item We asked respondents for their general views on blogs and blog posts, whereas we could have provided specific examples of blog posts and asked respondents about those examples. A different survey design might have revealed different criteria, however the current analyses and the synthesised checklists indicate that a rich set of criteria have been developed.
    
    \item We could have conducted more detailed follow--ups with both non--respondents and respondents. We intend to do this with those respondents who have expressed a willingness for a follow--up.
    
    \item A more sophisticated survey could ask multiple questions (phrased differently) about each credibility criteria so as to establish with confidence consistent responses. One explanation for the above limitations is that we wanted to conduct a survey that could be completed relatively quickly so as to reduce drop--out rates and increase the number of completed responses. 
    
    \item There is uncertainty as to what cases we should exclude and also uncertainty over what cases we should retain for the analyses. We have three different types of indicator:
    \begin{itemize}
        \item At least one respondent who completed the survey very quickly and who also responded with the same values (i.e., scored five for all nine criteria).
        \item One respondent who scored all Likert criteria consistently low.
        \item A small number of respondents who state in their qualitative comments that they do read blog posts, or do not read them often.
    \end{itemize}
    
\end{enumerate}

\subsection{Further research}

There are a number of directions for further research in this area, and again we enumerate them.
\begin{enumerate}
    \item An obvious direction for further research is to conduct more studies that address the threats and limitations identified in subsection \ref{subsection:threats}.
    
    \item With the raw responses from this survey available online, there is the opportunity for others to independently analyse the published data from the current study. 
    
    \item Similarly, there is the opportunity for others to independently replicate the existing study.
    
    \item The preliminary checklists warrant application, evaluation and revision, including extension. 
    
    \item It would also be valuable to widen the scope of this study to consider other kinds of grey literature and to compare the design, collection and analyses of grey literature in comparison to white literature. Such comparisons are timely as the research community increasingly uses social media in its research.
    
\end{enumerate}

\subsection{Conclusions}

Blog posts are a potential source of evidence for software engineering research however there are concerns relating to the quality of blog posts. We conducted a survey to better understand researchers' attitudes to blog posts. We explicitly asked respondents about nine pre--defined credibility criteria, as well as inviting respondents to identify additional criteria. The nine criteria were identified from an extensive review of prior literature on credibility. We found no clear majority consensus on which of the nine criteria to use to evaluate the credibility of blog posts. Instead, there appeared to be subgroups of respondents who broadly agreed on particular criteria.  We conducted thematic analyses of the qualitative comments from respondents to identify additional criteria. We then developed a preliminary suite of checklists for evaluating credibility, comprising several categories of criteria: foundational, process, author, content, media, influencer and reader. The checklists were synthesised from four sources of information: respondents' attitudes to the nine criteria, additional criteria proposed by respondents, criteria identified from the literature reviewed, and two quality checklists, one generic checklist for MLRs and one checklist for a specific GLR. We partially demonstrated the checklists by applying them back to the survey itself, and in so doing we exposed similarities between survey responses and blog posts. Finally, we reflected on implications for research and practice, threats to the validity of our work, and opportunities for further research.

\section*{Acknowledgements}
We thank the 44 survey respondents for their engagement in the survey. We thank members of Software Innovation NZ for their support in the development of the survey. We thank the reviewers in anticipation of their reviews. The survey was approved by the appropriate University of Canterbury (New Zealand) Ethics Committee (HEC 2017/68/LR-PS).

\bibliographystyle{plainnat} 
\bibliography{bibliography}

\begin{thebibliography}{46}
\providecommand{\natexlab}[1]{#1}
\providecommand{\url}[1]{\texttt{#1}}
\expandafter\ifx\csname urlstyle\endcsname\relax
  \providecommand{\doi}[1]{doi: #1}\else
  \providecommand{\doi}{doi: \begingroup \urlstyle{rm}\Url}\fi

\bibitem[Aniche et~al.(2018)Aniche, Treude, Steinmacher, Wiese, Pinto, Storey,
  and Gerosa]{aniche2018modern}
Maur{\'i}cio Aniche, Christoph Treude, Igor Steinmacher, Igor Wiese, Gustavo
  Pinto, Margaret-Anne Storey, and Marco~Aur{\'e}lio Gerosa.
\newblock How modern news aggregators help development communities shape and
  share knowledge.
\newblock In \emph{Int’l. Conf. on Software Engineering}, 2018.

\bibitem[Burton et~al.(2009)Burton, Java, Soboroff, et~al.]{burton2009icwsm}
Kevin Burton, Akshay Java, Ian Soboroff, et~al.
\newblock The icwsm 2009 spinn3r dataset.
\newblock In \emph{Third Annual Conference on Weblogs and Social Media (ICWSM
  2009)}, 2009.

\bibitem[Chau et~al.(2009)Chau, Xu, Cao, Lam, and Shiu]{chau2009blog}
Michael Chau, Jennifer Xu, Jinwei Cao, Porsche Lam, and Boby Shiu.
\newblock A blog mining framework.
\newblock \emph{It Professional}, 11\penalty0 (1):\penalty0 36--41, 2009.

\bibitem[Clarke and O’Connor(2012)]{clarke2012situational}
Paul Clarke and Rory~V O’Connor.
\newblock The situational factors that affect the software development process:
  Towards a comprehensive reference framework.
\newblock \emph{Information and Software Technology}, 54\penalty0 (5):\penalty0
  433--447, 2012.

\bibitem[Devanbu et~al.(2016)Devanbu, Zimmermann, and Bird]{devanbu2016belief}
Prem Devanbu, Thomas Zimmermann, and Christian Bird.
\newblock Belief \& evidence in empirical software engineering.
\newblock In \emph{Proceedings of the 38th international conference on software
  engineering}, pages 108--119. ACM, 2016.

\bibitem[Dyba(2013)]{dyba2013contextualizing}
Tore Dyba.
\newblock Contextualizing empirical evidence.
\newblock \emph{IEEE software}, 30\penalty0 (1):\penalty0 81--83, 2013.

\bibitem[Dyb{\aa} et~al.(2005)Dyb{\aa}, Kitchenham, and
  Jorgensen]{dyba2005evidence}
Tore Dyb{\aa}, Barbara~A Kitchenham, and Magne Jorgensen.
\newblock Evidence--based software engineering for practitioners.
\newblock \emph{IEEE software}, 22\penalty0 (1):\penalty0 58--65, 2005.

\bibitem[Dyb{\aa} et~al.(2012)Dyb{\aa}, Sj{\o}berg, and Cruzes]{dybaa2012works}
Tore Dyb{\aa}, Dag~IK Sj{\o}berg, and Daniela~S Cruzes.
\newblock What works for whom, where, when, and why?: on the role of context in
  empirical software engineering.
\newblock In \emph{Proceedings of the ACM--IEEE international symposium on
  Empirical software engineering and measurement}, pages 19--28. ACM, 2012.

\bibitem[Fenton et~al.(1994)Fenton, Pfleeger, and Glass]{fenton1994science}
Norman Fenton, Shari~Lawrence Pfleeger, and Robert~L. Glass.
\newblock Science and substance: A challenge to software engineers.
\newblock \emph{IEEE software}, 11\penalty0 (4):\penalty0 86--95, 1994.

\bibitem[Garousi and M{\"a}ntyl{\"a}(2016)]{garousi2016and}
Vahid Garousi and Mika~V M{\"a}ntyl{\"a}.
\newblock When and what to automate in software testing? a multi--vocal
  literature review.
\newblock \emph{Information and Software Technology}, 76:\penalty0 92--117,
  2016.

\bibitem[Garousi et~al.(2018)Garousi, Felderer, and
  M{\"a}ntyl{\"a}]{garousi2018guidelines}
Vahid Garousi, Michael Felderer, and Mika~V M{\"a}ntyl{\"a}.
\newblock Guidelines for including grey literature and conducting multivocal
  literature reviews in software engineering.
\newblock \emph{Information and Software Technology}, 2018.

\bibitem[Gordon and Swanson(2009)]{gordon2009identifying}
Andrew Gordon and Reid Swanson.
\newblock Identifying personal stories in millions of weblog entries.
\newblock In \emph{Third International Conference on Weblogs and Social Media,
  Data Challenge Workshop, San Jose, CA}, volume~46, 2009.

\bibitem[Inui et~al.(2008)Inui, Abe, Hara, Morita, Sao, Eguchi, Sumida,
  Murakami, and Matsuyoshi]{inui2008experience}
Kentaro Inui, Shuya Abe, Kazuo Hara, Hiraku Morita, Chitose Sao, Megumi Eguchi,
  Asuka Sumida, Koji Murakami, and Suguru Matsuyoshi.
\newblock Experience mining: Building a large--scale database of personal
  experiences and opinions from web documents.
\newblock In \emph{Proceedings of the 2008 IEEE/WIC/ACM International
  Conference on Web Intelligence and Intelligent Agent Technology--Volume 01},
  pages 314--321. IEEE Computer Society, 2008.

\bibitem[Ivarsson and Gorschek(2011)]{Ivarsson2011}
Martin Ivarsson and Tony Gorschek.
\newblock A method for evaluating rigor and industrial relevance of technology
  evaluations.
\newblock \emph{Empirical Software Engineering}, 16\penalty0 (3):\penalty0
  365--395, Jun 2011.
\newblock ISSN 1573--7616.
\newblock \doi{10.1007/s10664--010--9146--4}.
\newblock URL \url{https://doi.org/10.1007/s10664-010-9146-4}.

\bibitem[Khan et~al.(2017)Khan, Daud, Ishfaq, Amjad, Aljohani, Abbasi, and
  Alowibdi]{khan2017modelling}
Hikmat~Ullah Khan, Ali Daud, Umer Ishfaq, Tehmina Amjad, Naif Aljohani,
  Rabeeh~Ayyaz Abbasi, and Jalal~S Alowibdi.
\newblock Modelling to identify influential bloggers in the blogosphere: A
  survey.
\newblock \emph{Computers in Human Behavior}, 68:\penalty0 64--82, 2017.

\bibitem[Kitchenham and Charters(2007)]{Kitchenham07guidelinesforSLRs}
B.~Kitchenham and S~Charters.
\newblock Guidelines for performing systematic literature reviews in software
  engineering, 2007.

\bibitem[Kitchenham and Pfleeger(2002)]{kitchenham2002surveypart5}
Barbara Kitchenham and Shari~Lawrence Pfleeger.
\newblock Principles of survey research: part 5: populations and samples.
\newblock \emph{ACM SIGSOFT Software Engineering Notes}, 27\penalty0
  (5):\penalty0 17--20, 2002.

\bibitem[Kitchenham et~al.(2002)Kitchenham, Pfleeger, Pickard, Jones, Hoaglin,
  El~Emam, and Rosenberg]{kitchenham2002preliminary}
Barbara~A Kitchenham, Shari~Lawrence Pfleeger, Lesley~M Pickard, Peter~W Jones,
  David~C. Hoaglin, Khaled El~Emam, and Jarrett Rosenberg.
\newblock Preliminary guidelines for empirical research in software
  engineering.
\newblock \emph{IEEE Transactions on software engineering}, 28\penalty0
  (8):\penalty0 721--734, 2002.

\bibitem[Kurashima et~al.(2006)Kurashima, Tezuka, and
  Tanaka]{kurashima2006mining}
Takeshi Kurashima, Taro Tezuka, and Katsumi Tanaka.
\newblock Mining and visualizing local experiences from blog entries.
\newblock In \emph{DEXA}, pages 213--222. Springer, 2006.

\bibitem[Kurashima et~al.(2009)Kurashima, Fujimura, and
  Okuda]{kurashima2009discovering}
Takeshi Kurashima, Ko~Fujimura, and Hidenori Okuda.
\newblock Discovering association rules on experiences from large--scale blog
  entries.
\newblock In \emph{European Conference on Information Retrieval}, pages
  546--553. Springer, 2009.

\bibitem[Lakshmanan and Oberhofer(2010)]{lakshmanan2010knowledge}
Geetika Lakshmanan and Martin Oberhofer.
\newblock Knowledge discovery in the blogosphere: Approaches and challenges.
\newblock \emph{IEEE internet computing}, 14\penalty0 (2):\penalty0 24--32,
  2010.

\bibitem[Pagano and Maalej(2011)]{pagano2011developers}
Dennis Pagano and Walid Maalej.
\newblock How do developers blog?: an exploratory study.
\newblock In \emph{Proceedings of the 8th working conference on Mining software
  repositories}, pages 123--132. ACM, 2011.

\bibitem[Park et~al.(2010)Park, Jeong, and Myaeng]{park2010detecting}
Keun~Chan Park, Yoonjae Jeong, and Sung~Hyon Myaeng.
\newblock Detecting experiences from weblogs.
\newblock In \emph{Proceedings of the 48th Annual Meeting of the Association
  for Computational Linguistics}, pages 1464--1472. Association for
  Computational Linguistics, 2010.

\bibitem[Parnin and Treude(2011)]{parnin2011measuring}
Chris Parnin and Christoph Treude.
\newblock Measuring api documentation on the web.
\newblock In \emph{Proceedings of the 2nd international workshop on Web 2.0 for
  software engineering}, pages 25--30. ACM, 2011.

\bibitem[Parnin et~al.(2013)Parnin, Treude, and Storey]{parnin2013blogging}
Chris Parnin, Christoph Treude, and Margaret-Anne Storey.
\newblock Blogging developer knowledge: Motivations, challenges, and future
  directions.
\newblock In \emph{Program Comprehension (ICPC), 2013 IEEE 21st International
  Conference on}, pages 211--214. IEEE, 2013.

\bibitem[Petersen and Gencel(2013)]{petersen2013worldviews}
Kai Petersen and Cigdem Gencel.
\newblock Worldviews, research methods, and their relationship to validity in
  empirical software engineering research.
\newblock In \emph{2013 Joint Conference of the 23rd International Workshop on
  Software Measurement and the 8th International Conference on Software Process
  and Product Measurement}, pages 81--89. IEEE, 2013.

\bibitem[Petersen and Wohlin(2009)]{petersen2009context}
Kai Petersen and Claes Wohlin.
\newblock Context in industrial software engineering research.
\newblock In \emph{Proceedings of the 2009 3rd International Symposium on
  Empirical Software Engineering and Measurement}, pages 401--404. IEEE
  Computer Society, 2009.

\bibitem[Petersen et~al.(2015)Petersen, Vakkalanka, and
  Kuzniarz]{petersen2015guidelines}
Kai Petersen, Sairam Vakkalanka, and Ludwik Kuzniarz.
\newblock Guidelines for conducting systematic mapping studies in software
  engineering: An update.
\newblock \emph{Information and Software Technology}, 64:\penalty0 1--18, 2015.

\bibitem[Prechelt and Petre(2010)]{prechelt2010credibility}
Lutz Prechelt and Marian Petre.
\newblock Credibility, or why should i insist on being convinced?
\newblock \emph{Making Software: What Really Works, and Why We Believe It},
  page~17, 2010.

\bibitem[Rainer(2017)]{rainer2017using}
Austen Rainer.
\newblock Using argumentation theory to analyse software practitioners’
  defeasible evidence, inference and belief.
\newblock \emph{Information and Software Technology}, 2017.

\bibitem[Rainer and Williams(2018)]{rainerASWEC2018}
Austen Rainer and Ashley Williams.
\newblock Using blog articles in software engineering research: benefits,
  challenges and case--survey method.
\newblock In \emph{Proceedings of the 25th Australasian Software Engineering
  Conference (ASWEC 2018)}, 2018.

\bibitem[Rainer and Williams(2019)]{rainer2019using}
Austen Rainer and Ashley Williams.
\newblock Using blog-like documents to investigate software practice: Benefits,
  challenges, and research directions.
\newblock \emph{Journal of Software: Evolution and Process}, 31\penalty0
  (11):\penalty0 e2197, 2019.

\bibitem[Rainer et~al.(2003)Rainer, Hall, and Baddoo]{rainer2003persuading}
Austen Rainer, Tracy Hall, and Nathan Baddoo.
\newblock Persuading developers to" buy into" software process improvement: a
  local opinion and empirical evidence.
\newblock In \emph{Empirical Software Engineering, 2003. ISESE 2003.
  Proceedings. 2003 International Symposium on}, pages 326--335. IEEE, 2003.

\bibitem[Raulamo-Jurvanen et~al.(2017)Raulamo-Jurvanen, M\"{a}ntyl\"{a}, and
  Garousi]{RaulamoJurvanen2017choosing}
P\"{a}ivi Raulamo-Jurvanen, Mika M\"{a}ntyl\"{a}, and Vahid Garousi.
\newblock Choosing the right test automation tool: A grey literature review of
  practitioner sources.
\newblock In \emph{Proceedings of the 21st International Conference on
  Evaluation and Assessment in Software Engineering}, EASE'17, pages 21--30,
  New York, NY, USA, 2017. ACM.
\newblock ISBN 978--1--4503--4804--1.
\newblock \doi{10.1145/3084226.3084252}.
\newblock URL \url{http://doi.acm.org/10.1145/3084226.3084252}.

\bibitem[Runeson and H{\"o}st(2009)]{runeson2009guidelines}
Per Runeson and Martin H{\"o}st.
\newblock Guidelines for conducting and reporting case study research in
  software engineering.
\newblock \emph{Empirical software engineering}, 14\penalty0 (2):\penalty0 131,
  2009.

\bibitem[Runeson et~al.(2012)Runeson, H{\"o}st, Rainer, and
  Regnell]{runeson2012case}
Per Runeson, Martin H{\"o}st, Austen Rainer, and Bjorn Regnell.
\newblock \emph{Case study research in software engineering: Guidelines and
  examples}.
\newblock John Wiley \& Sons, 2012.

\bibitem[Soldani et~al.(2018)Soldani, Tamburri, and Van
  Den~Heuvel]{soldani2018pains}
Jacopo Soldani, Damian~Andrew Tamburri, and Willem-Jan Van Den~Heuvel.
\newblock The pains and gains of microservices: A systematic grey literature
  review.
\newblock \emph{Journal of Systems and Software}, 2018.

\bibitem[Storey et~al.(2010)Storey, Treude, van Deursen, and
  Cheng]{storey2010impact}
Margaret-Anne Storey, Christoph Treude, Arie van Deursen, and Li-Te Cheng.
\newblock The impact of social media on software engineering practices and
  tools.
\newblock In \emph{Proceedings of the FSE/SDP workshop on Future of software
  engineering research}, pages 359--364. ACM, 2010.

\bibitem[Storey et~al.(2014)Storey, Singer, Cleary, Figueira~Filho, and
  Zagalsky]{storey2014r}
Margaret-Anne Storey, Leif Singer, Brendan Cleary, Fernando Figueira~Filho, and
  Alexey Zagalsky.
\newblock The (r) evolution of social media in software engineering.
\newblock In \emph{Proceedings of the on Future of Software Engineering}, pages
  100--116. ACM, 2014.

\bibitem[Swanson et~al.(2014)Swanson, Rahimtoroghi, Corcoran, and
  Walker]{swanson2014identifying}
Reid Swanson, Elahe Rahimtoroghi, Thomas Corcoran, and Marilyn Walker.
\newblock Identifying narrative clause types in personal stories.
\newblock In \emph{Proceedings of the 15th Annual Meeting of the Special
  Interest Group on Discourse and Dialogue (SIGDIAL)}, pages 171--180, 2014.

\bibitem[Williams(2018{\natexlab{a}})]{williams2018software}
Ashley Williams.
\newblock Do software engineering practitioners cite research on software
  testing in their online articles?: A preliminary survey.
\newblock In \emph{Proceedings of the 22nd International Conference on
  Evaluation and Assessment in Software Engineering 2018}, pages 151--156. ACM,
  2018{\natexlab{a}}.

\bibitem[Williams(2018{\natexlab{b}})]{williams2018using}
Ashley Williams.
\newblock Using reasoning markers to select the more rigorous software
  practitioners' online content when searching for grey literature.
\newblock In \emph{Proceedings of the 22nd International Conference on
  Evaluation and Assessment in Software Engineering 2018}, pages 46--56. ACM,
  2018{\natexlab{b}}.

\bibitem[Williams and Rainer(2018)]{williams_rainer_cred_tr}
Ashley Williams and Austen Rainer.
\newblock \emph{The analysis and synthesis of previous work on credibility
  assessment in online media: technical report}.
\newblock University of Canterbury, New Zealand, 04 2018.
\newblock URL
  \url{https://www.researchgate.net/publication/324765770_The_analysis_and_synthesis_of_previous_work_on_credibility_assessment_in_online_media_technical_report}.

\bibitem[Williams and
  Rainer(2019{\natexlab{a}})]{williamsEASE19credibilitysurvey}
Ashley Williams and Austen Rainer.
\newblock How do software engineering researchers assess the credibility of
  practitioner--generated blog posts?
\newblock In \emph{23rd International Conference on Evaluation and Assessment
  in Software Engineering (EASE'19), 14 -- 17 April 2019, Copenhagen, Denmark},
  2019{\natexlab{a}}.

\bibitem[Williams and
  Rainer(2019{\natexlab{b}})]{williamsEASE19practitionerscite}
Ashley Williams and Austen Rainer.
\newblock Do software engineering practitioners cite research in their online
  articles? a larger scale replication.
\newblock In \emph{23rd International Conference on Evaluation and Assessment
  in Software Engineering (EASE'19), 14 -- 17 April 2019, Copenhagen, Denmark},
  2019{\natexlab{b}}.

\bibitem[Wohlin(2013)]{wohlin2013evidence}
Claes Wohlin.
\newblock An evidence profile for software engineering research and practice.
\newblock In \emph{Perspectives on the Future of Software Engineering}, pages
  145--157. Springer, 2013.

\end{thebibliography}

\section*{Appendix}

\subsection*{Appendix A - Survey questions}

Table \ref{table:survey-questions} presents the questions used in the survey together with permissible values. For Q2, we used a text field to allow respondents flexibility in response. For Q3, the range of Likert values was \textit{O: No blog is credible} to \textit{5: All blogs are credible}. For the Q5 sub--questions, the range of Likert values was \textit{0: Not at all important} to \textit{6: Extremely important}.

\begin{table}[!htbp]
    \centering
    \small
    \caption{Summary of survey questions}
    \label{table:survey-questions}
    \begin{tabular}{ p{0.5cm}  p{13.5cm}  p{1cm} } 
    % \hline
    \textbf{\#}	& \textbf{Question} & \textbf{Values}\\
    \hline
    Q1	&	Please summarise your specific areas of research? Please be more specific than 'software engineering' :). Separate each area with a semi-colon e.g. software testing; requirements engineering.	&	Text	\\
    Q2	&	How many years experience do you have conducting software engineering research? (Enter number to one decimal place.)	&	Text	\\
    Q3	&	In general, how credible do you consider blog articles that are written by practitioners?	&	Likert\\
    Q4	&	Please provide general comments to complement your rating.  (Optional.)	&	Text	\\
    Q5.1	&	For each of the following criteria, please rate how important you consider the criterion to be when assessing quality: - The clarity of writing within the blog article	&	Likert\\
    Q5.2	&	For each of the following criteria, please rate how important you consider the criterion to be when assessing quality: - The reporting of empirical data within the blog article that relates to the claims made in the article	&	Likert\\
    Q5.3	&	For each of the following criteria, please rate how important you consider the criterion to be when assessing quality: - The reporting of the method by which the empirical data was collected and analysed	&	Likert\\
    Q5.4	&	For each of the following criteria, please rate how important you consider the criterion to be when assessing quality: - The reporting of professional experience within the blog article that relates to the claims made in the blog article	&	Likert\\
    Q5.5	&	For each of the following criteria, please rate how important you consider the criterion to be when assessing quality: - The presence of web links to other practitioner sources that relates to the claims made in the article	&	Likert\\
    Q5.6	&	For each of the following criteria, please rate how important you consider the criterion to be when assessing quality: - The presence of web links to peer-reviewed research that relates to the claims made in the article	&	Likert\\
    Q5.7	&	For each of the following criteria, please rate how important you consider the criterion to be when assessing quality: - The presence of reasoning within the blog that relates to the claims made in the article	&	Likert\\
    Q5.8	&	For each of the following criteria, please rate how important you consider the criterion to be when assessing quality: - The prior beliefs of the reader when reading the article	&	Likert\\
    Q5.9	&	For each of the following criteria, please rate how important you consider the criterion to be when assessing quality: - The influence of other people's opinions on the reader's beliefs e.g. a recommendation to read the blog article	&	Likert\\
    Q6	&	Please provide more information to explain your answers.	&	Text	\\
    Q7	&	Are there any other criteria that you think are missing or have not been considered by this survey? Please summarise those other criteria.	&	Text	\\
    Q8	&	Do you think that the criteria identified in Q5 and Q7 generalise to assessing the quality of content written by practitioners, other than blogs (e.g. emails, Q\&A sites such as Stack Exchange, comments that have been provided in response to blog articles)?	&	Text	\\
    Q9	&	Please provide more information to explain your answer	&	Text	\\
    Q10	&	Do you think the criteria identified in Q5 and Q7 generalise to assessing the quality of content written by researchers e.g. journal articles, conference papers?	&	Text	\\
    Q11	&	Please provide more information to explain your answer.	&	Text	\\
    Q12	&	Please provide any other comments you would like to make on this survey.	&	Text	\\

    \hline
    \end{tabular}
\end{table}

\subsection*{Appendix B - Bubble plots for the criteria within the survey}
\begin{figure}
    \centering
    \subfloat[Empirical data]{\includegraphics[scale=0.2]{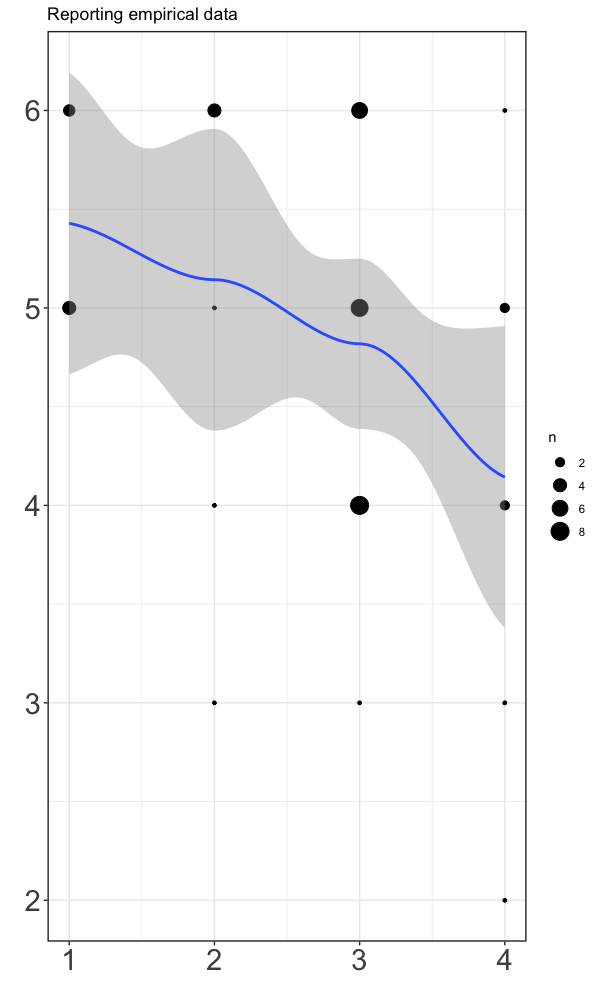}\label{subfigure:RED-bubble}}
    \subfloat[Reasoning]{\includegraphics[scale=0.2]{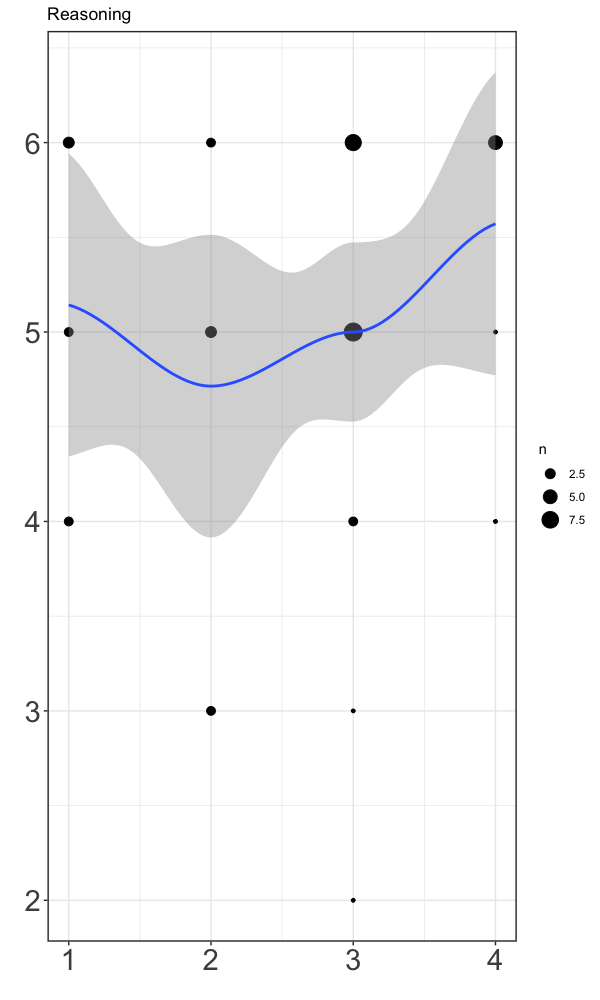}}\\
    \subfloat[Research method]{\includegraphics[scale=0.2]{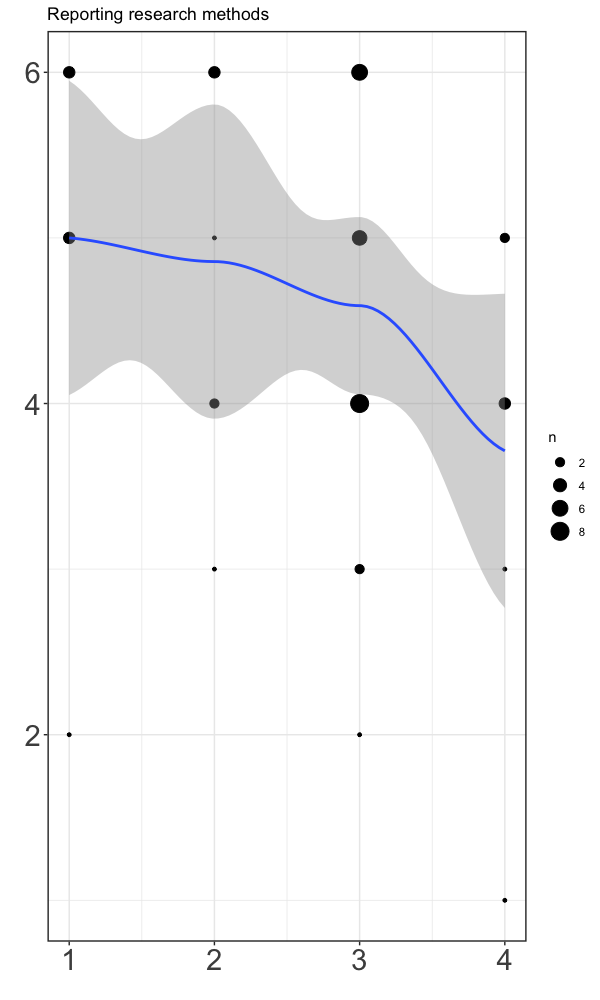}\label{subfigure:RM-bubble}}
    \subfloat[Clarity of writing]{\includegraphics[scale=0.2]{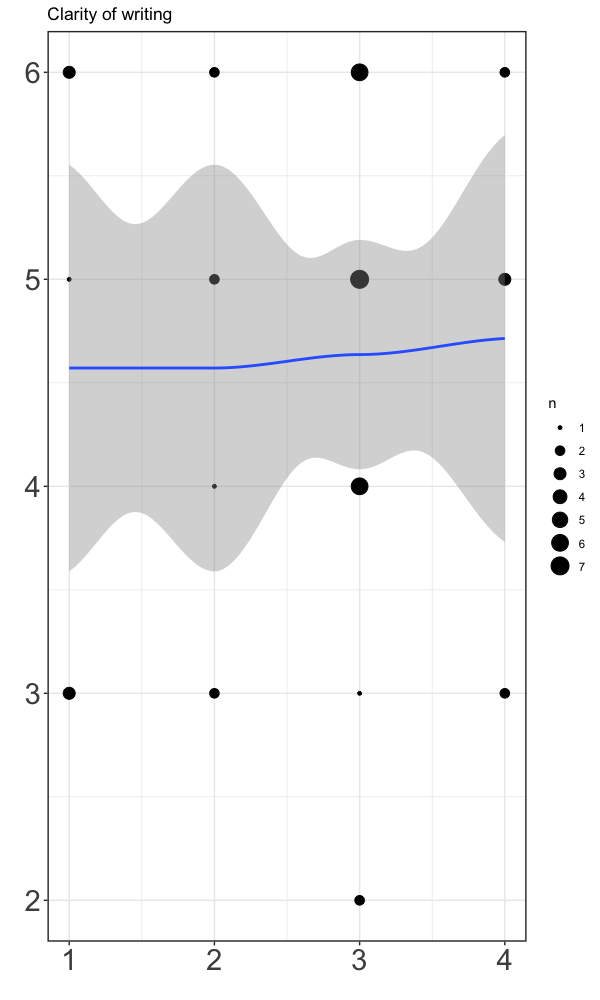}}\\    
    \caption{Bubble charts with Loess smoother and confidence interval}
    \label{figure:RED-RM-Rsn-scatter-plots}
\end{figure}

\begin{figure}
    \centering
    \subfloat[Citing research]{\includegraphics[scale=0.2]{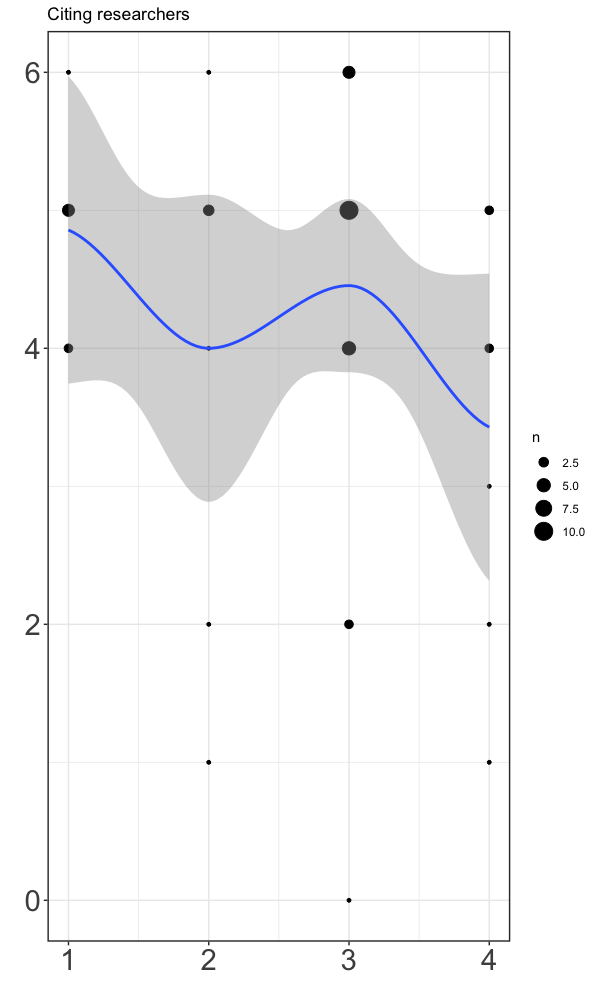}}
    \subfloat[Citing practice]{\includegraphics[scale=0.2]{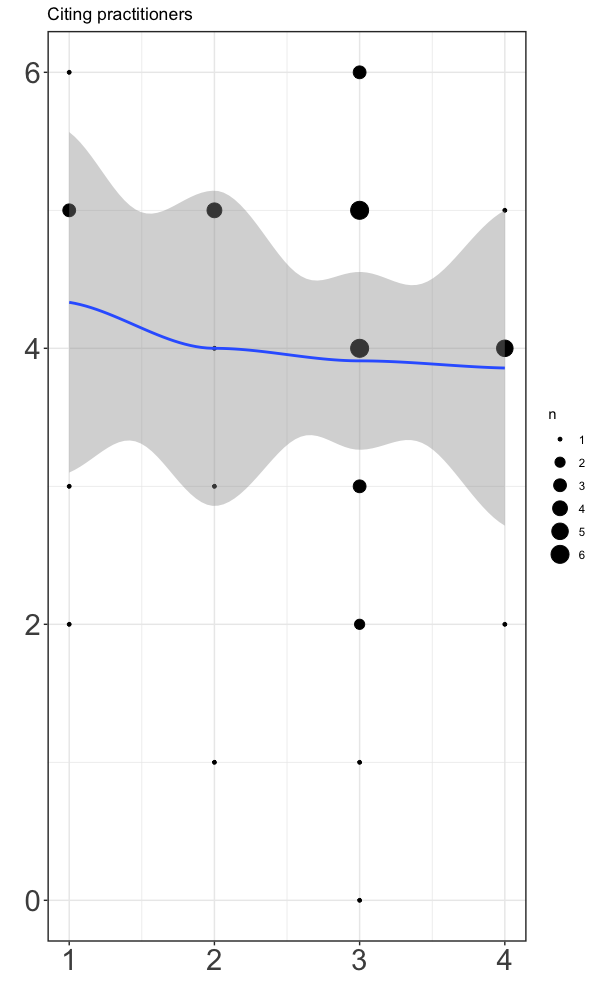}}\\ 
    \caption{Dotplots with Loess smoother and confidence interval}
    \label{figure:CoW-URLR-URLP-scatter-plots}
\end{figure}

\begin{figure}
    \centering
    \subfloat[Prof. experience]{\includegraphics[scale=0.2]{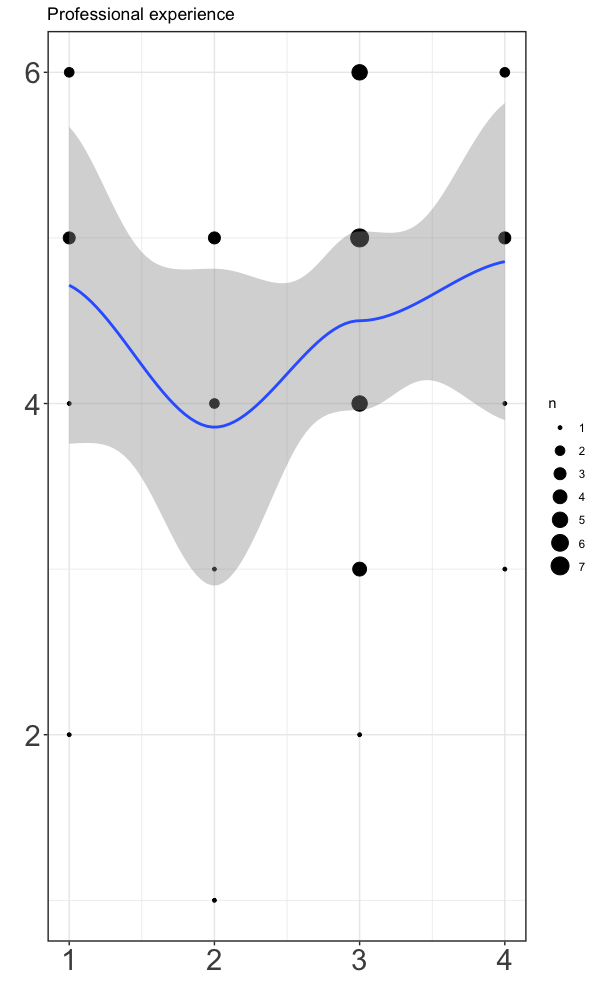}}
    \subfloat[Others' influence]{\includegraphics[scale=0.2]{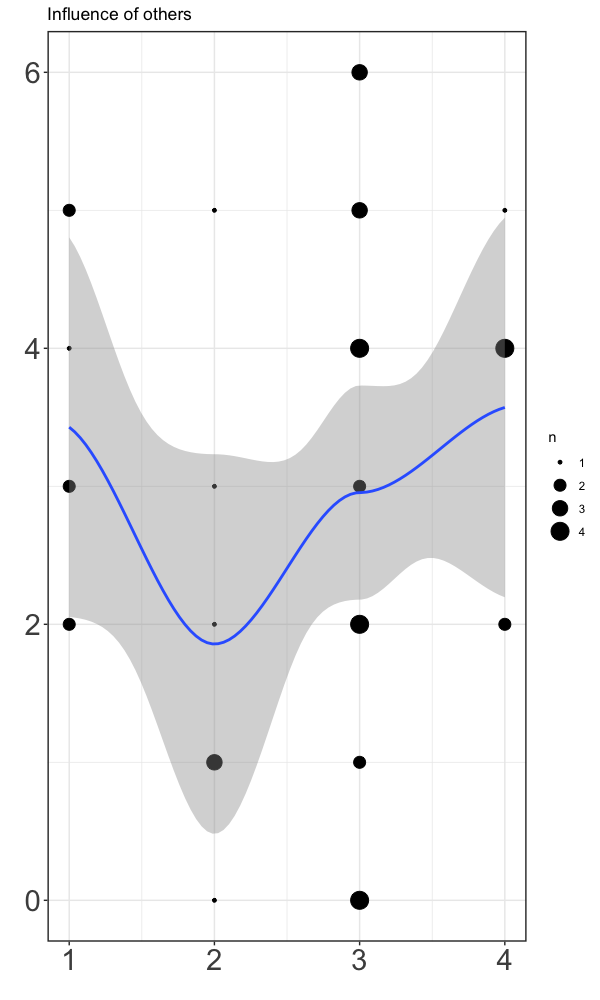}}
    \subfloat[Prior beliefs]{\includegraphics[scale=0.2]{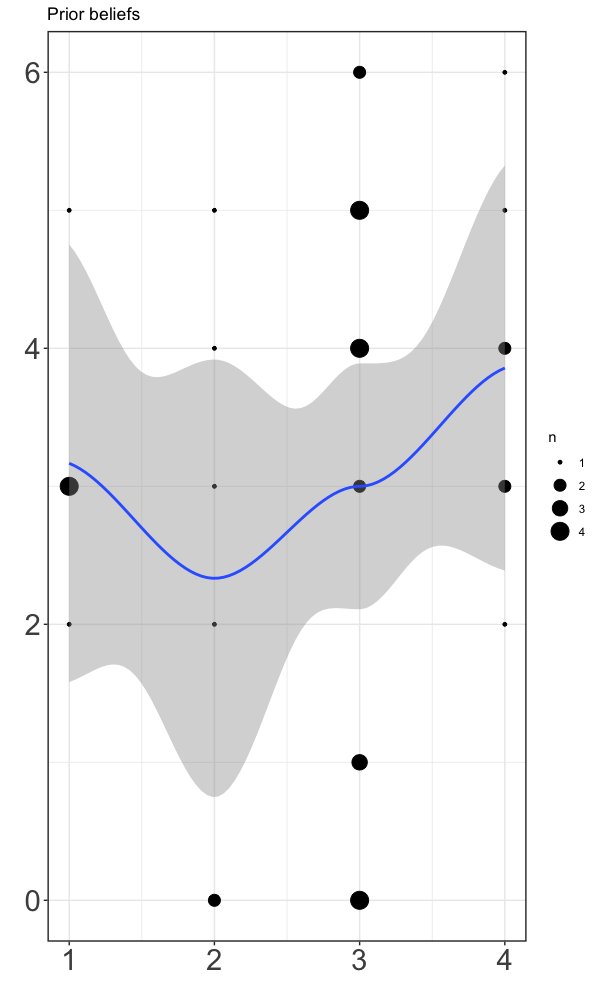}}\\
    \caption{Dotplots with Loess smoother and confidence interval}
    \label{figure:Pexp-Others-Beliefs-scatter-plots}
\end{figure}

Figures \ref{figure:RED-RM-Rsn-scatter-plots}, \ref{figure:CoW-URLR-URLP-scatter-plots}, and \ref{figure:Pexp-Others-Beliefs-scatter-plots} present bubble plots for each of the nine criteria and, in so doing, the plots provide a kind of transposition of the data presented in the heatmap. 

For each bubble plot the \textit{x}--axis presents the scores for the General Credibility (see \textbf{GC} in Table \ref{figure:heatmap-of-responses}), the \textit{y}--axis presents the scores for the respective criterion (in the range 0 -- 6), and the size of the dot indicates the number of responses. A LOESS (locally estimated scatterplot smoothing) line is plotted, together with (again) a confidence interval. Once again, we use these plots for the purpose of exploratory data analyses. And once again, we proceed cautiously in the interpretation of the data, e.g., the dataset is clearly imbalanced with about half of the responses plotted at $x = 3$. 

Figure \ref{subfigure:RED-bubble} and Figure \ref{subfigure:RM-bubble} both suggest, unsurprisingly, an inverse relationship between the credibility of blogs posts and the importance of reporting empirical data and reporting research methods: respondents who place a higher value on reporting empirical data and on reporting research methods also place a lower value on the credibility of blog posts, and vice versa. As noted earlier, this is unsurprising given the value that researchers place on evidence and on evidence--collection. In our excluded--case analysis, respondent \#22 appears to take a `purest' view of this relationship, i.e., rejecting entirely the other seven criteria.

For the three sub--figures in Figure \ref{figure:Pexp-Others-Beliefs-scatter-plots}, there is curious behaviour where $x = 1$ and where $x = 2$, i.e., when the score for General Credibility is 1 or 2. In all three figures, it is not clear whether the ratings for the criterion are unusually high for $x = 1$ or unusually low for $x = 2$; or whether instead the data is dominated by responses at $x = 3$. And these patterns suggest the presence of sub--populations.

\end{document}